\documentclass[useAMS,usenatbib]{mn2e}

\usepackage{amssymb, amsmath, epsfig, bm, url, fixltx2e}
\newcommand{\bfn}{{\boldsymbol{n}}}
\newcommand{\bfk}{{\boldsymbol{k}}}

\newcommand{\bfr}{{\boldsymbol{r}}}
\newcommand{\bfC}{{\boldsymbol{C}}}
\newcommand{\bfQ}{{\boldsymbol{Q}}}
\newcommand{\bfX}{{\boldsymbol{X}}}
\newcommand{\matF}{{\boldsymbol{F}}}
\newcommand{\bfx}{{\boldsymbol{x}}}

\newcommand{\HeII}{He{\sc ~ii}}

\begin{document}

\title{On Estimating Ly$\alpha$ Forest Correlations between Multiple Sightlines}

\author[M. McQuinn and M. White]{Matthew McQuinn\thanks{mcquinn@uw.edu}$^{1}$ and Martin White$^{1,2}$\\ \\
$^{1}$ Department of Astronomy, University of California, Berkeley, CA 94720\\
$^{2}$ Department of Physics, University of California, Berkeley, CA 94720\\}

\pubyear{2011} \volume{415} \pagerange{2257-2269}

\maketitle\label{firstpage}

\begin{abstract}
The next frontier of Ly$\alpha$ forest studies is the reconstruction of $3$D correlations from a dense sample of background sources.  The measurement of $3$D correlations has the potential to improve constraints on fundamental cosmological parameters, ionizing background models, and the reionization history.  This study addresses the sensitivity of spectroscopic surveys to $3$D correlations in the Ly$\alpha$ forest.  We show that the sensitivity of a survey to this signal can be quantified by just a single number, a noise-weighted number density of sources on the sky.  We investigate how the sensitivity of a spectroscopic quasar (or galaxy) survey scales as a function of its depth, area, and redshift.  We propose a simple method for weighting sightlines with varying $S/N$ levels to estimate the correlation function, and we show that this estimator generally performs nearly as well as the minimum variance quadratic estimator.  In addition, we show that the sensitivity of a quasar survey to the flux correlation function is generally maximized if it observes each field just long enough to achieve $S/N\approx2$ in a $1~$\AA\ pixel on an $L_*$ quasar while acquiring spectra for all quasars with $L > L_*$:  Little is gained by integrating longer on the same targets or by including fainter quasars.  We quantify how these considerations relate to constraints on the angular diameter distance, the curvature of space-time, and the reionization history.
\end{abstract}


\section{introduction}

The Ly$\alpha$ forest is an established tool for studying structure formation, the intergalactic medium (IGM), and cosmological parameters.  It has been used to place constraints on the linear-theory density correlations at $\sim 1~$comoving Mpc separations, smaller separations than other large-scale structure probes \citep{1997ApJ...490..564W, 1998ApJ...495...44C, 2000ApJ...543....1M, 2005PhRvD..71j3515S, 2005PhRvD..71f3534V}.  In addition, it is our best tool for studying the thermal history of the intergalactic gas \citep{schaye00, mcdonald01b, lidz09, becker10}, and it has been used to place a lower bound on the redshift of reionization \citep{2001AJ....122.2850B, fan06}.  However, past Ly$\alpha$ forest analyses have only utilized correlations in the transmission within a single sightline (e.g., \citealt{mcdonald05b}) or between a small number of sightlines (e.g., \citealt{2000ApJ...532...77W, 2000MNRAS.311..657L, hennawi07}).  This approach was justified because of the low sky density of quasars in previous widefield surveys.  However, the next generation of surveys will achieve densities of $\approx 10-100$ quasars per deg$^2$, allowing $S/N>1$ on tens of comoving Mpc $3$D modes.  The ongoing Baryon Oscillation Spectroscopic Survey (BOSS) on the $2.5\,$m Sloan Telescope aims to measure $3$D correlations from $1.6\times 10^5$ quasars over $8000~$deg$^2$.   A similar survey but on a $4\,$m telescope, BigBOSS, is anticipated to begin in 2016.\footnote{\url{http://cosmology.lbl.gov/BOSS/}, \url{http://bigboss.lbl.gov/index.html}}  In general, piggybacking a $3$D Ly$\alpha$ survey on an optical spectroscopic survey comes with little cost owing to the low sky density of $z>2$ quasars.

The advantages of $3$D Ly$\alpha$ forest correlations have been enumerated in several previous theoretical studies.  The traditional approach utilizing purely line-of-sight measurements will always be more sensitive to parameters that impact correlations on less than several Mpc scales.  However, unlike with line-of-sight measurements, with $3$D correlation measurements it is possible to detect the baryon acoustic oscillation feature at high precision and to use it as a standard ruler \citep{white03, mcdonald07, slosar09, white10}.  Such a measurement at $z=2-3$ with a BOSS-like spectroscopic quasar survey has the potential to rule out early dark energy models and to place $10^{-3}$--level constraints on $\Omega_k$, the $z=0$ spatial curvature density \citep{mcdonald07}.  Three-dimensional measurements are also more sensitive to large-scale intensity and temperature fluctuations than are line-of-sight ones \citep{white10, mcquinn10}.  In fact, the resulting temperature fluctuations from  physically motivated models for \HeII\ reionization can change the Ly$\alpha$ forest correlation function by $\mathcal{O}(1)$ at $100~$Mpc separations.  This difference would be easily detected in a measurement utilizing correlations between sightlines, but would have gone undetected in previous line-of-sight analyses \citep{mcquinn10}.  In addition, a future Ly$\alpha$ forest survey that is sensitive to $z\approx4$ could detect the relic temperature fluctuations from hydrogen reionization \citep{mcquinn10}. 

This paper studies the ability of spectroscopic quasar surveys to constrain correlations in the $3$D Ly$\alpha$ forest.  Section \ref{sec:sensitivity} presents formulae for the sensitivity of spectroscopic surveys to $3$D correlations as well as for how to weight sightlines based on their signal-to-noise properties.  We use the resulting expressions to understand the sensitivity of both purely hypothetical and proposed quasar surveys as a function of depth, volume, and redshift.  Section \ref{sec:QE} discusses how the suggested weights relate to the minimum variance quadratic estimator.
Section \ref{sec:considerations} discusses additional survey considerations:  (1) whether taking the spectra of high-redshift galaxies in addition to quasars can enhance the sensitivity, (2) the advantages of cross correlating a $3$D Ly$\alpha$ survey with another survey of large-scale structure, (3) how to optimize a survey's strategy to minimize the uncertainty in its estimate of the correlation function, and (4) how the previous considerations translate to constraints on cosmological parameters and interesting astrophysical processes.  Finally, Section \ref{sec:systematics} discusses the severity of continuum subtraction errors, of errors in the mean flux estimate, and of contamination from damping wings.  We find that these systematics are likely to be less severe for upcoming $3$D Ly$\alpha$ forest analyses than they were for past line-of-sight studies.

We assume a flat $\Lambda$CDM cosmological model where necessary with $\Omega_m=0.27$, $h=0.71$, $\sigma_8= 0.8$, $n_s=0.96$, and $\Omega_b = 0.046$ \citep{komatsu10}.  We henceforth will use ``Mpc'' as shorthand for ``comoving Mpc,''  and we use the standard Fourier convention for cosmological studies in which $2 \, \pi$'s appear under the $dk$'s.  Table~1 provides relevant numbers for the Ly$\alpha$ forest that are used in our calculations.

\section{Sensitivity to $3$D Flux Power Spectrum}
\label{sec:sensitivity}

\begin{table}
\caption{Relevant numbers and conversion factors:  $\chi$ is the conformal distance, $\Delta \chi$ is the conformal distance covered by the Ly$\alpha$ forest that is not contaminated by the Ly$\beta$ forest, and $b$ is the large-scale bias of the Ly$\alpha$ forest in our model.  The mean flux, $\langle F \rangle$, is calculated using the fitting formula in \citet{meiksin09}, and $d\lambda/d\chi$ and $dv/d\chi$ are in units of \AA\ Mpc$^{-1}$ and km~s$^{-1}$~Mpc$^{-1}$, respectively.}
\begin{center}
\begin{tabular}{c c c c c c c}
\hline \hline
  
  $z$ &   $\lambda$ [\AA] & $\langle F \rangle$ & $\Delta \chi$ [Mpc] &  $d\lambda/d\chi$ &  $dv/d\chi$ & $|b|$\\ \hline
  $2.0$  &  $3647$   & $0.88$  & $782$   &    $0.82$    &     $67$ & $0.12$\\
  $2.5$  &  $4255$   &  $0.80$ & $740$    &   $1.01$    &     $71$ & $0.18$\\
  $3.0$  &  $4863$   &  $0.70$  & $701$    &   $1.22$    &     $75$ & $0.27$\\
  $3.5$  &  $5471$   & $0.58$  & $666$    &   $1.45$    &    $79$ & $0.37$\\
  $4.0$  &  $6078$  &   $0.44$  & $635$    &   $1.69$    &     $83$ & $0.55$\\
\hline
\end{tabular}
\end{center}
\label{table:simple}\label{table:conversions}
\end{table}

\subsection{Covariance}
\label{ss:covariance}

Let us take a quasar survey that provides $N$ Ly$\alpha$ forest spectra at redshift $z$.  It provides these Ly$\alpha$ forest spectra towards the locations $\bfx_{\perp, n}$ on the sky, where $n \in \{1, ..., N\}$.  We treat the line-of-sight direction as continuous for convenience such that the survey measures the Ly$\alpha$ forest in a spatial window given by
\begin{equation}
W(\bfx) = \bar{n}^{-1} \, \sum_{n=1}^{N} \delta^D \, \left(\bfx_\perp - \bfx_{\perp,n} \right),
\end{equation}
where $\bar{n} \equiv N/{\cal A}$, ${\cal A}$ is the survey area on the sky, and $\bfx = (x_{\parallel}, \bfx_\perp)$.  We use an analogous convention for the Fourier wavevector, $\bfk$.  Thus, the survey measures ${\delta(\bfx)} = \delta_{\rm F}(\bfx) \, W(\bfx)$, where $\delta_{\rm F}(\bfx) = F/\langle F \rangle - 1$ is the overdensity in the transmission fraction $F$, where $\langle F \rangle$ is the mean transmission.   Switching to the Fourier basis, the covariance of two $\delta$-modes is
\begin{eqnarray}
\langle \tilde{\delta}(\bfk) \tilde{\delta}({\bfk'})^{*} \rangle =  \bar{n}^{-2}\,(2 \pi)\, \delta^D(k_{\parallel} - k_{\parallel}') 
\label{eqn:cov}
\end{eqnarray}
\begin{eqnarray}
& &\times~  \sum_{n=1}^{N}\sum_{m=1}^{N} \tilde{w}_n (k_{\parallel}) \tilde{w}_m(k'_{\parallel})^* e^{-i (\bfk_\perp \cdot \bfx_{\perp, n} - \bfk_\perp' \cdot \bfx_{\perp, m})}  \nonumber \\
& &\times ~\left[P_{{\rm N},n} \, \delta_{n}^m + \int \frac{d^2 \bfk_\perp^{''}}{(2\pi)^2} e^{i \bfk_\perp^{''} \cdot (\bfx_{\perp, n} - \bfx_{\perp, m})}  P_{\rm F}(k_{\parallel}, \bfk_\perp^{''})  \right] \nonumber,
\end{eqnarray}
where $\langle ... \rangle$ represents an ensemble average, $P_{{\rm N}, n}$ is the $1$D noise power for skewer $n$ (which we assume does not correlate with other sightlines), $\delta_{n}^m$ is the Kronecker delta, tildes signify the Fourier basis, and $\tilde{w}_n (k_{\parallel})$ is the weight given to the mode for skewer $n$ and normalized such that $N^{-1} \, \sum_n \tilde{w}_n (k_{\parallel}) = 1$.  We will often keep the $k_{\parallel}$ dependence of $\tilde{w}_n$ implicit for notational simplicity.  This weighting scheme assumes that the combined weight of pixels $n$ and $m$ factorizes.  While not fully general, with the proper choice of $w_n$, this scheme is close to optimal (Section \ref{sec:QE}).  Lastly, equation (\ref{eqn:cov}) approximates the modes as continuous in the transverse direction, as would occur in the limit ${\cal A} \rightarrow \infty$.    

The noise power that appears in equation (\ref{eqn:cov}) can be related to the $[S/N]_{X}$ on the continuum in a pixel in a resolution element of size $X$:
\begin{eqnarray}
P_{{\rm N}, n} &=& \langle F \rangle^{-2} \, [S/N]_{\Delta x}^{-2} \left(\frac{\Delta x}{1 \, {\rm Mpc}} \right),\\
 &= & 0.8 \, \langle F \rangle^{-2} \, [S/N]_{\Delta \lambda}^{-2} \left(\frac{{\Delta \lambda}}{1 ~{\rm \AA}}\right) \left(\frac{1+z}{4}\right)^{-3/2},
\end{eqnarray}
where $\Delta \lambda$ is the pixel size in wavelength, and we have assumed that the noise is white.  We will henceforth discuss the noise in terms of the $S/N$ in $1~$\AA\ pixels.  However, for the tens of Mpc modes of interest for $3$D Ly$\alpha$ forest analyses, resolving $1\,$\AA\ may not be required.  A resolution of tens of \AA\ suffices to capture the scales where $3$D correlations can be detected for upcoming measurements \citep{mcdonald07}.  It is simply a convenience that leads us to quote the $S/N$ per $1~$\AA\ pixel.

Equation (\ref{eqn:cov}) is the covariance of the Fourier-space flux overdensity field, $\tilde{\delta}_{\rm F}$, convolved with the survey window, $\widetilde{W}$.  In the limit of large $N$, $\widetilde{W}$ becomes a $\delta$-function such that $\tilde{\delta} \approx \tilde{\delta}_{\rm F}$.  The covariance of two modes in the flux field can be approximated in this limit as diagonal with entries\footnote{The largest of the off-diagonal terms is $P_{\rm F} \, {\rm VAR}[\overline{{\tilde{w}^2}}/{N}]$, where the ``VAR'' function represents a number whose probability distribution function has a variance given by the argument.  Its exact value is determined by the $\bfx_{\perp, n}$.  Even though there are $N^2$ possible correlations, the error from neglecting the off-diagonal terms scales as $N^{-1/2}$ rather than $N^{-1}$ because there are only $N$ quasar locations.}
\begin{eqnarray}
{P}_{\rm tot} \equiv \langle |\tilde{\delta}_k|^2 \rangle = P_{\rm F}(\bfk) + \bar{n}^{-1} \, \left[ P_{\rm N} +  \overline{\tilde{w}^2} \,P_{\rm los}(k_{\parallel}) \right],\label{eqn:Ptot}
\end{eqnarray}
where $\overline{X}$ indicates the average of quantity $X_n$ over $N$ skewers, 
\begin{equation}
P_{\rm los}(k_{\parallel}) \equiv \int \frac{d^2 \bfk_\perp}{(2 \pi)^2}\, P_{\rm F}(k_{\parallel}, \bfk_\perp),
\label{eqn:Plos}
\end{equation}
and
\begin{equation}
P_{\rm N} \equiv \frac{1}{N} \, \sum_{n}^N |\tilde{w}_n|^2 P_{{\rm N}, n}.
\end{equation}
The function $P_{\rm los}$ is just the line-of-sight power spectrum that is typically measured in cosmological studies of the Ly$\alpha$ forest (e.g., \citealt{mcdonald05b}).  As in \citet{mcdonald07}, we will refer to this term in equation (\ref{eqn:Plos}) as the ``aliasing'' term.  Equation~(\ref{eqn:Ptot}) is derived by noting that the sum over $N$ skewers is a monte-carlo evaluation of the integral, and the monte-carlo integral of a plane wave leads to a $\delta$-function to precision of $N^{-1/2}$.  Replacing these summations with $\delta$-functions results in a two $\delta$-function term from the components that have $m\neq n$ (that yields $P_{\rm F}$) and a one $\delta$-function term from $m = n$ (that yields $P_{\rm tot}-P_{\rm F}$).  Equation~(\ref{eqn:Ptot}) agrees with equation~(13) in \citet{mcdonald07} and with equation~(A6) in \citet{white10}.\footnote{As a consistency check, it is simple to show that $\langle \chi^2(k_{\parallel}) \rangle \equiv \sum P_{\rm F}/P_{\rm tot}$ is $\leq N$, where the sum is over all wavevectors with the same $k_{\parallel}$.  This inequality demonstrates that the amount of information that can be extracted from $\tilde{\delta}_F$ (as measured by the significance at which $\tilde{\delta}_F$ can be detected) is less than or equal to the number of Ly$\alpha$ forest pixels in a survey.}

Henceforth we ignore correlations between flux pixels in a single line-of-sight, which contain a small fraction of the total information.  This omission is equivalent to enforcing the condition that $m \neq n$ in the summation in equation~(\ref{eqn:cov}).  We define the estimator $\widehat{P}_{\rm F} \equiv  |\tilde{\delta}_\bfk|^2$ for this case.  Dropping terms suppressed by higher powers of $N$, the ensemble average of $\widehat{P}_{\rm F}$ is
\begin{equation}
\langle \widehat{P}_{\rm F} (\bfk) \rangle = P_{\rm F}(\bfk).\label{eqn:PF}
\end{equation}
Thus, omitting line-of-sight correlations results in an unbiased estimator at lowest order.  While $\widehat{P}_{\rm F}$ is biased to fractional order $N^{-1/2}$, the noise in this estimate can be corrected because the $\bfx_{\perp,n}$ are known as shown in Section \ref{sec:QE}.  An additional advantage of omitting line-of-sight correlations is that the contamination from continuum fluctuations and damping wings primarily induce this type of correlation (Section \ref{sec:systematics}).

The covariance of $\widehat{P}_{\rm F}(\bfk)$ is
\begin{eqnarray}
 {\rm cov}[\widehat{P}_{\rm F}(\bfk), \; \widehat{P}_{\rm F}(\bfk')]_{k_{\parallel}} = {\cal A}^2 \, \bar{n}^{-4} \int \frac{d^2 \bfk_\perp^{(1)} \, d^2 \bfk_\perp^{(2)}}{(2\pi)^4}~ \sum_{n, m, k, l}^{n\neq m, k \neq l}~~~~ \nonumber
\end{eqnarray}
\begin{eqnarray}
&\times& e^{i \bfx_n \cdot (\bfk_\perp - \bfk_\perp^{(1)})} e^{-i \bfx_m \cdot ( \bfk_\perp - \bfk_\perp^{(2)})} \, \bigg[e^{i \bfx_k \cdot (\bfk_\perp' - \bfk_\perp^{(2)})} \nonumber \\
&&\times ~e^{-i \bfx_l \cdot (\bfk_\perp' - \bfk_\perp^{(1)})} + e^{i \bfx_k \cdot (\bfk_\perp' + \bfk_\perp^{(2)})} e^{-i \bfx_l \cdot (\bfk_\perp' + \bfk_\perp^{(1)})} \bigg] \nonumber \\
& \times&  \tilde{w}_n \, \tilde{w}_l^* \, \left[P_{\rm F}(k_{\parallel}, \bfk_\perp^{(1)}) + P_{{\rm N}, n} \, \delta_{n}^l  \, (2\pi)^2 \,  \delta^D(\bfk_\perp^{(1)})\right] \nonumber \\  
&\times &  \tilde{w}_k \, \tilde{w}_m^* \, \left[P_{\rm F}(k_{\parallel}, \bfk_\perp^{(2)}) + P_{{\rm N}, m} \,\delta_{m}^k \,(2\pi)^2 \,\delta^D(\bfk_\perp^{(2)}) \right], \label{eqn:covP}
\end{eqnarray}
where all $3$D wavevectors that appear are evaluated at $k_{\parallel}$, and we have assumed Gaussianity and, thus, dropped the connected $4$th moment.  While the small-scale modes in the forest are far from being Gaussian, Gaussianity is likely to be a decent approximation at the $> 10~$Mpc wavelength modes of interest for $3$D analyses.  With a bit of algebra analogous to how equation (\ref{eqn:Ptot}) was derived, this equation simplifies to
\begin{eqnarray}
  &{\rm cov}[\widehat{P}_{\rm F}(\bfk), \widehat{P}_{\rm F}(\bfk')]_{k_{\parallel}} = 2 P_{\rm tot}^2 \delta_{\bfk}^{\bfk'}   + \frac{4}{N}\, \overline{\tilde{w}^2} \, P_{\rm F}(\bfk) P_{\rm F}(\bfk') ~~~   \label{eqn:covPk}\\
                & +  \frac{1}{N}\, \overline{\tilde{w}^2}^2 \,  \bar{n}^{-1}  \, \int \frac{d^2 \bfk_\perp^{(1)}}{(2\pi)^2} \, P_{\rm F}(k_{\rm ll}, \bfk_\perp^{(1)})  P_{\rm F}(k_{\rm ll}, \bfk_\perp^{(1)} -\bfk_\perp - \bfk_\perp') \nonumber \\
                                & +  \frac{1}{N}\, \overline{\tilde{w}^2}^2 \,  \bar{n}^{-1}  \, \int \frac{d^2 \bfk_\perp^{(1)}}{(2\pi)^2} \, P_{\rm F}(k_{\rm ll}, \bfk_\perp^{(1)})  P_{\rm F}(k_{\rm ll}, \bfk_\perp^{(1)} + \bfk_\perp - \bfk_\perp'). \nonumber
\end{eqnarray}

For the wavevectors and quasar number densities of interest, the off-diagonal terms in equation (\ref{eqn:covPk}) are unimportant.  The subtlety here
is that, unlike the diagonal terms, these terms do not average down linearly with the number of independent modes when binning the estimates for $P_{\rm F}$ in a shell in $k$-space.  This is
analogous to the well-known behavior of the $4$-point function in surveys
of large-scale structure \citep{meiksin99}, which has become known as
``beat-coupling'' \citep{HRS06,RimHam06}.  The most important off-diagonal contribution
comes from $4 \, N^{-1} \,P_{\rm F}(\bfk)P_{\rm F}(\bfk')$.
At fixed $k_{||}$, this becomes comparable to the corresponding diagonal term when
the number of binned pixels in the shell is comparable to $N$.  This criterion is satisfied for larger wavevectors than
$k_\perp=0.06 \, [\bar{n}/10^{-3}\,{\rm Mpc}^{-2}]^{1/2}\times\sqrt{k_\perp/\Delta k_\perp}\,{\rm Mpc}^{-1}$,
where $\Delta k_\perp$ is the size of the bin in the $k_\perp$ direction.
However, for the wavevectors and number densities of relevance, we find that in practice this term is always subdominant to the aliasing term, $\bar n^{-1}\, P_{\rm F}$, in $P_{\rm tot}$.  The inclusion of these terms is however
formally important in that they serve to ``cap'' the total information content
of a survey, resolving an issue raised in \citet{mcdonald07}.  
The off-diagonal elements are a more important consideration when line-of-sight correlations are included.

Quasar clustering was not accounted for in the previous expressions.  While clustering does not bias $\widehat{P}_F$ when line-of-sight correlations are omitted (at least at quadratic order in the density), it does increase the variance of the estimate.
Clustering results in the term $\bar n^{-1} \, (P_{\rm N} + P_{\rm los})$ in $P_{\rm tot}$ gaining the factor $1+ C_q(\bfk_\perp) \, \bar{n}$, where $C_q(\bfk_\perp)$ is the angular power spectrum of the sources that contribute at that redshift.  This correction is significant on scales where the clustering power is comparable or larger than the shot power.   Assuming a quasar bias of $3$ at $z=3$ and that each Ly$\alpha$ spectrum yields $500~$Mpc of absorption, the two become comparable for $\bar{n} \approx 10^{-2}~$Mpc$^{-2}$ (and then only at $k_\perp \approx 0.02~$Mpc$^{-1}$, the horizon scale at matter-radiation equality).  This $\bar n$ is an order of magnitude higher than what upcoming surveys will achieve.

\subsection{Optimal Weights}
\label{ss:weights}

\begin{table}
\caption{Sloan Digital Sky Survey constraints on $P_{\rm los}$ in units of Mpc from \citet{mcdonald05b}.}
\begin{center}
\begin{tabular}{c | c c c c c}
\hline \hline
$k_\parallel$   &     $z=2.2$&     $z=2.6$&     $z=3.0$&     $z=3.6$&     $z=4.0$\\ \hline  
0.15 & $0.27(2)$ 
& $0.43(2)$
& $0.58(4)$
& $1.05(10)$
& $1.47(22)$ \\  
   
0.20 & $0.26(1)$
& $0.38(2)$
& $0.58(3)$
& $0.86(6)$
& $1.08(13)$\\

0.30 & $0.18(1)$
& $0.30(1)$
& $0.44(2)$
& $0.81(5)$
& $0.85(10)$\\

0.50 & $0.15(1)$
& $0.24(1)$
& $0.35(1)$
& $0.59(3)$
& $0.81(07)$\\ \hline 
\end{tabular}
\end{center}
\label{table:Plos}
\end{table}

\begin{table}
\caption{Value of $\nu_n \equiv (1 + P_{{\rm N}, n}/P_{\rm los})^{-1}$ (c.f., eqn. \ref{eqn:neff}) at several redshifts as a function of the $S/N$ on the continuum in a $1~$\AA\ pixel.  This calculation uses the estimate of \citet{mcdonald05b} at $k = 0.0014~$s~km$^{-1}$ for $P_{\rm los}$.}
\begin{center}
\begin{tabular}{c | c c c c c}
\hline \hline
$[S/N]_{1A}$ & $z=2.2$ & $z=2.6$ & $z=3.0$ & $z=3.6$ & $z=4.0$\\ \hline
0.50  & 0.04 & 0.06 & 0.08 & 0.10 & 0.11 \\
1.00  & 0.14 & 0.20 & 0.27 & 0.31 & 0.33 \\
2.00  & 0.40 & 0.50 & 0.59 & 0.64 & 0.66 \\
5.00  & 0.81 & 0.86 & 0.90 & 0.92 & 0.93 \\
10.00  & 0.94 & 0.96 & 0.97 & 0.98 & 0.98\\ \hline
\end{tabular}
\end{center}
\label{table:SNR}
\end{table}

The weight that maximizes the signal-to-noise ratio, $P_{\rm F}^2/{\rm var}[\widehat{P}_{\rm F}]$, is
\begin{equation}
\tilde{w}_n(k_{\parallel}) = \mathcal{B}(k_{\parallel}) \, \left(P_{\rm los}(k_{\parallel}) + P_{{\rm N}, n} \right)^{-1},
\label{eqn:wn}
\end{equation}
where $\mathcal{B}(k_{\parallel})$ is fixed by our normalization convention and absorbs all factors that do not depend on $n$, and this maximization neglects the subdominant off-diagonal covariances so that ${\rm var}[\widehat{P}_{\rm F}] = 2 \, P_{\rm tot}^2$.  (We show in Section \ref{ss:realspace} that a simple generalization of these weights can be easily applied to real data.)  This choice of weights results in ${P}_{\rm tot}$ becoming
\begin{equation}
{P}_{\rm tot}(\bfk) = P_{\rm F}(\bfk) + \bar{n}_{\rm eff}^{-1} \,P_{\rm los}(k_{\parallel}), 
\label{eqn:Pkwithw}
\end{equation}
where
\begin{equation}
\bar{n}_{\rm eff} \equiv \frac{1}{\cal A}  \, \sum_{n=1}^{N} \nu_n,  ~~~ \nu_n \equiv  \frac{P_{\rm los}(k_{\parallel}) }{ P_{\rm los}(k_{\parallel}) + P_{{\rm N}, n}}.
\label{eqn:neff}
\end{equation}
Thus, a single number, $\bar{n}_{\rm eff}$, characterizes the sensitivity of a Ly$\alpha$ forest survey to $P_{\rm F}$, and $\nu_n$ is a measure of the importance of each quasar on a scale of $0$ to $1$.  While $\bar{n}_{\rm eff}$ depends on $k_\parallel$, in practice this dependence is likely to be weak because $P_{\rm los}(k_{\parallel})$ has roughly a white noise power spectrum at relevant wavevectors and $\bar{n}_{\rm eff}$ also depends relatively weakly on $P_{\rm los}(k_{\parallel})$.  The constancy of $P_{\rm los}$ at $k \leq 0.5~$Mpc$^{-1}$ is quantified in Table \ref{table:Plos}, which tabulates measurements of $P_{\rm los}$ from \citet{mcdonald05b} at several $k_{\parallel}$ and redshifts.  Constancy should be an even better approximation at smaller $k_{\parallel}$ than is tabulated.  For a BOSS-like Ly$\alpha$ forest survey at $z=2.5$, a factor of $2$ smaller $P_{\rm los}(k_{\parallel})$ from its small-$k_{\parallel}$ asymptote results in only a factor of $1.4$ decrease in $\bar{n}_{\rm eff}$.  The decrease is even smaller for a deeper survey.

The gains in sensitivity to $P_{\rm F}$ are meager from improving the $S/N$ on a quasar once $P_{{\rm N}, n} < P_{\rm los}$, which corresponds roughly to $[S/N]_{1A} \approx 2$.   Table \ref{table:SNR} quantifies this statement by giving $\nu_n$ as a function of the $S/N$ at $1~$\AA, using the estimates of $P_{\rm los}$ at the smallest $k$ quoted in \citet{mcdonald05b} of $0.0014~$s~km$^{-1}$.  However, these numbers should be applicable over a wide range of $k$ owing to the form of $P_{\rm los}$.  Even though the amount of power in the forest increases with redshift, the $S/N$ requirements at fixed source flux remain nearly constant with increasing redshift (becoming slightly less stringent).

\begin{table}
\caption{$\bar{n}_{\rm eff}$ for a quasar survey in units of $10^{-3}~$Mpc$^{-2}$ assuming that $m_{\rm AB}^{\rm lim} = m_{\rm AB}^{1A}$ and $k = 0.1~$Mpc$^{-1}$.  These numbers are calculated using the B-band luminosity function from \citet{hopkins06}.  The second number in select entries is this but also using galaxies in addition to quasars.  The bottom row provides the factor that converts $\bar{n}_{\rm eff}$ in each respective column to units of deg$^{-2}$.}
\begin{center}
\begin{tabular}{c | c c c c c }
\hline \hline
$m_{\rm AB}^{1A}$ & $z=2$ & $z=2.5$ & $z=3$ & $z=4$ & $z=5$ \\ \hline
21  & 0.26 & 0.16 & 0.07 & 0.01 & 0.00 \\
22  & 0.82 & 0.54 & 0.27 & 0.04 & 0.00 \\
23  & 2.0 & 1.4 & 0.78 & 0.11 & 0.01 \\
24  & 3.8/5 & 2.8 & 1.7 & 0.32 & 0.04 \\
25  & 6.4/18 & 4.8/9 & 3.2/5 & 0.77 & 0.10 \\ 
26  & 9.8/66 & 7.4/47 & 5.1/28 & 1.5/4 & 0.24 \\ \hline
 & 8.4 & 11 & 13 & 16 & 19\\
\hline
\end{tabular}
\end{center}
\label{table:neff}
\end{table}

\begin{figure}
\epsfig{file=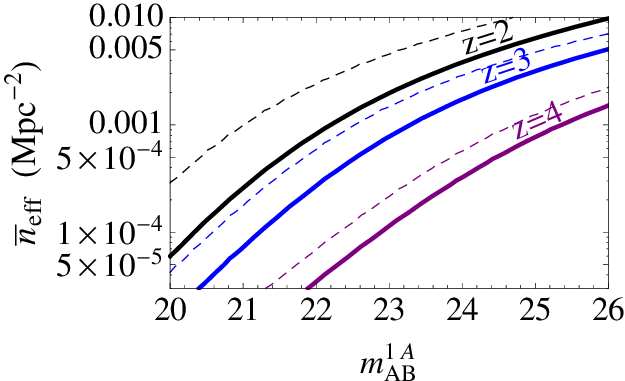, width=8cm}
\caption{Solid curves show the effective number density of quasars contributing Ly$\alpha$ forest spectra at redshift $z$ (eqn. \ref{eqn:neff}) as a function of the B-band AB magnitude that has $[S/N]_{1A}=1$, $m_{\rm AB}^{1A}$.  We have also assumed that $m_{\rm AB}^{1A}$ is equal to the limiting magnitude of the survey.  The dashed curves are the actual number of quasars brighter than $m_{\rm AB}$.   These curves are calculated for $k_{\parallel} = 0.1~$Mpc$^{-1}$ using the \citet{hopkins06} luminosity function and assuming that $S/N \propto\,$flux. \label{fig:Neff}}
\end{figure}

Figure \ref{fig:Neff} shows $\bar{n}_{\rm eff}$ at three redshifts as a function of the B-band magnitude at which a survey obtains $[S/N]_{1\, A} = 1$.  For this and subsequent calculations, we assume that $P_{\rm F}$ has the simple form
\begin{equation}
P_{\rm F}(\bfk, z) = b(z)^2 \,(1 + g \, \mu^2)^2 \; P_\delta^{\rm lin}(k, z) \; \exp[- k_{\parallel}^2/k_{\rm D}^2],
\end{equation}
where $P_\delta^{\rm lin}$ is the linear-theory density power spectrum and $k_{\rm D}^2 \equiv m_p/[k_{\rm B} T]$.  We assume $T = 20,000~$K isothermal gas such that $k_{\rm D} = 0.08$~s$^{-1}$~km, and $b(z)$ is calibrated so that $P_{\rm los}$ matches the \citet{mcdonald05b} measurements at $0.0014~$s$^{-1}$~km (see Table~1).  The factor $(1 + g \, \mu^2)^2$ owes to peculiar velocities, where $\mu = \hat{\bfn} \cdot \hat{\bfk}$ and $\hat{\bfn}$ is the unit vector along the line-of-sight.  We set $g=1$, a choice motivated by the results of \citet{slosar09} and \citet{mcquinn10}, but caution that there is at present theoretical uncertainty in this choice at the $50\%$ level \citep{mcquinn10}. 

The solid curves in Figure \ref{fig:Neff} are calculated using the \citet{hopkins06} B-band luminosity function and assuming that each quasar contributes $\approx 500~$Mpc of Ly$\alpha$ absorption (specifically the forest between $1041$ and $1185~$\AA\ in each sightline).  Thus, these curves are the effective number density of quasars projected over this distance, which can be thought of as the number of quasars that contribute Ly$\alpha$ forest spectra at the stated redshift.   These curves also take the survey limiting magnitude $m^{\rm lim}_{\rm AB}$ to be equal to the magnitude where quasars have $[S/N]_{1A} = 1$, $m_{\rm AB}^{1A}$, a choice we discuss below, and assume that $S/N \propto\,$flux.  This scaling corresponds to the sensitivity being sky- or dark current-limited, a choice motivated by the fact that this scaling matters most for the faintest objects that are observed.  However, the value of $\bar{n}_{\rm eff}$ is not significantly changed if rather we assume $S/N \propto\,$flux$^{1/2}$ as would occur if the observations were photon-limited.  The solid curves from top to bottom are $\bar{n}_{\rm eff}$ for $z=2, ~3,$ and $4$.   These curves illustrate that it will be challenging to (1) measure the $3$D Ly$\alpha$ forest at $z> 3$ because of the falloff in the total abundance of quasars or (2) obtain $\bar{n}_{\rm eff} \gg 10^{-3}~$Mpc$^{-2}$ at any redshift because of the shallow faint-end slope of the luminosity function.  Table \ref{table:neff} tabulates $\bar{n}_{\rm eff}$ as various redshifts and limiting magnitudes, and it also gives the conversion from Mpc$^{-2}$ to deg$^{-2}$ units.

The dashed curves in Figure \ref{fig:Neff} show the total number of quasars brighter than $m_{\rm AB}$.  The effective number density at $m_{\rm AB}^{1A}$ is always a factor of a few smaller than the total number of quasars, with the difference decreasing with redshift.  See Appendix \ref{ap:powerlaw} for an analytic understanding of this suppression for power-law luminosity functions.

 BOSS will achieve a B-band magnitude limit of $m^{\rm lim}_{\rm AB} \approx 22$ at $z=2$ ($m^{\rm lim}_{\rm AB} \approx 21$ at $z=3$) and also achieve $m_{\rm AB}^{1A} \approx 22$.  These numbers yield $\bar{n}_{\rm eff} = (0.3 -0.8) \times 10^{-3}~$Mpc$^{-2}$ [$4-7~$deg$^{-2}$] at $z=2-3$.   BigBOSS aims to achieve one magnitude fainter than BOSS, which results in $\bar{n}_{\rm eff} = (1 -2) \times 10^{-3}~$Mpc$^{-2}$ at $z=2-3$.  For these $\bar{n}_{\rm eff}$ and at $k \gtrsim 0.1~$Mpc$^{-1}$, $S/N$ scales linearly with $\bar{n}_{\rm eff}$ such that BigBOSS will be a few times more sensitive than BOSS.

\begin{figure}
\epsfig{file=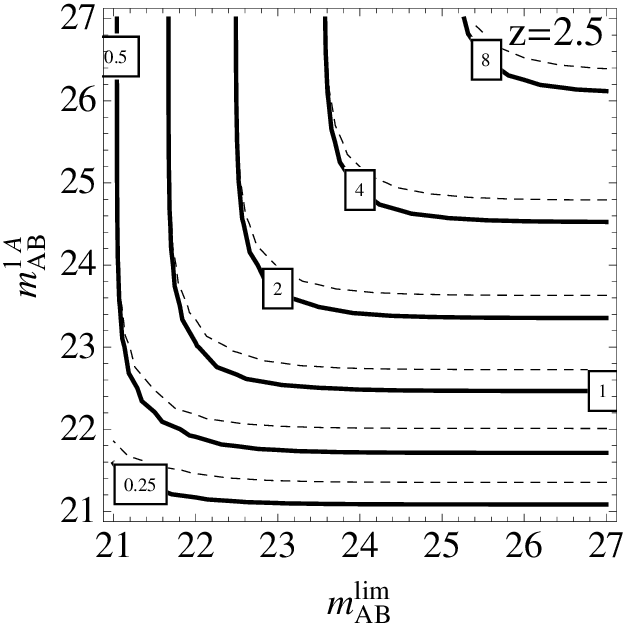, width=8cm}
\caption{Impact of varying the limiting magnitude of the survey, $m^{\rm lim}_{\rm AB}$, along with the magnitude at which $[S/N]_{1A} = 1$, $m_{\rm AB}^{1A}$.  The solid curves are contours of constant $\bar{n}_{\rm eff}$ evaluated at $k_\parallel = 0.1~$Mpc$^{-1}$ and $z=2.5$, with the labels in units of $10^{-3}~$Mpc$^{-2}$.  The corresponding dashed curves are the same but evaluated at  $k_\parallel = 0.5~$Mpc$^{-1}$.   \label{fig:Neffcont}}
\end{figure}

Figure \ref{fig:Neffcont} is a contour plot of $\bar{n}_{\rm eff}$ as a function of $m^{\rm lim}_{\rm AB}$ and $m_{\rm AB}^{1A}$. The solid contours are calculated assuming $k_{\parallel} = 0.1~$Mpc$^{-1}$ and are labeled in units of $10^{-3}~$Mpc$^{-2}$. These curves illustrate that $\bar{n}_{\rm eff}$ is maximized roughly when $m^{\rm lim}_{\rm AB} \approx m_{\rm AB}^{1A}-0.5$.  Little is gained by observing fainter quasars than $m_{\rm AB}^{1A} - 0.5$ or by integrating longer on the same quasars once $m_{\rm AB}^{1A} \approx m^{\rm lim}_{\rm AB} + 0.5$.  The dashed contours are the same but at $k_{\parallel} = 0.5~$Mpc$^{-1}$ (at which $P_{\rm los}$ is a factor of $0.6$ smaller in our model), demonstrating that the $k_{\parallel} $ dependence of these conclusions is weak.

How much is gained by weighting by $w_n$ relative to a uniform weighting scheme?  Let us take a luminosity with power-law index $-2$, a form discussed in Appendix \ref{ap:powerlaw}.  We assume that $S/N \propto {\rm flux}$, $P_{\rm N, lim}/P_{\rm los} = 1$ (equivalent to $[S/N]_{1A} \approx 2$; Table \ref{table:SNR}), where $P_{\rm N, lim}$ is the noise power spectrum for quasars at the survey limiting magnitude.  This case results in only a 4\% improvement in the value of $P_{\rm tot} - P_{\rm F}$ relative to a simple uniform weighting.  For $P_{\rm N, lim}/P_{\rm los} = 0.5 \,(0.3)$, the improvement is $40\%$ ($60\%$).  Thus, weighting only offers a significant improvement when a large fraction of the quasars have $P_{\rm N, n}/P_{\rm los} < 1$.

\subsection{Effective Volume}
\label{sec:veff}

\begin{figure}
\epsfig{file=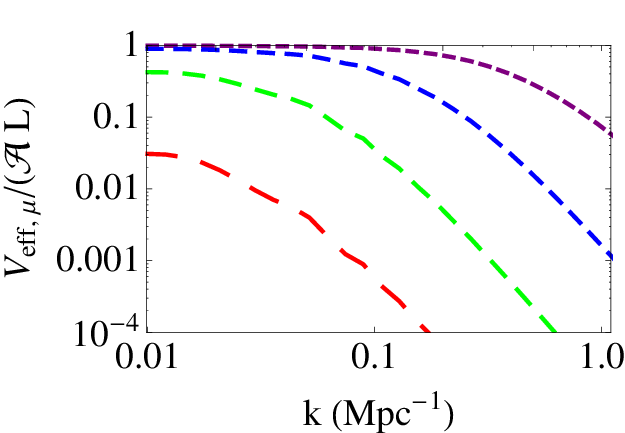, width=8cm}
\caption{$3$D Ly$\alpha$ forest effective volume divided by survey volume at $|\bfk|$, $z=3$, and for $\bar{n}_{\rm eff} = 10^{-1}, ~10^{-2}, ~10^{-3}$, and $10^{-4}~$Mpc$^{-2}$ in order of increasing dash length (decreasing amplitude). The plotted quantity is equal to $(2 \, P_{\rm F, \mu}/\delta P_{\rm F, \mu})^{2}$ and is essentially independent of redshift at the plotted scales. \label{fig:Veff}}  
\end{figure}

One can define an effective volume for Ly$\alpha$ forest surveys analogous to the effective volume in galaxy surveys \citep{feldman94}.  
In particular, in the galaxy survey case the effective volume is
\begin{equation}
V_{\rm eff,g}(\bfk) \equiv V_{\rm g} \, \left(\frac{P_{\rm g}(\bfk)}{P_{\rm g}(\bfk) + \bar{n}_{\rm g, 3D}^{-1}} \right)^2,\label{eqn:Veffg}
\end{equation}
which becomes in the Ly$\alpha$ forest survey case
\begin{equation}
V_{\rm eff}(\bfk) \equiv {\cal A} \, L \,  \left(\frac{P_{\rm F}(\bfk)}{P_{\rm F} (\bfk) + \bar{n}_{\rm eff}^{-1} P_{\rm los}(k_{\parallel})}\right)^2.
\label{eqn:Veff}
\end{equation}
In equation (\ref{eqn:Veffg}) and (\ref{eqn:Veff}), $V_{\rm g}$ is the actual volume of the galaxy survey, $P_{\rm g}$ is the galaxy power spectrum, $\bar{n}_{\rm g, 3D}$ is the number density of galaxies, and $L$ is the line-of-sight dimension of the Ly$\alpha$ survey.   The major difference between these two effective volumes is just that in the Ly$\alpha$ case the shot noise is in the plane of the sky and this term is modulated by the line-of-sight power.

Figure \ref{fig:Veff} plots a generalization of the the effective volume, $V_{\rm eff, \mu}(k)/[{\cal A} \, L]$, for $\bar{n}_{\rm eff} = 10^{-1}, ~10^{-2}, ~10^{-3}$, and $10^{-4}~$Mpc$^{-2}$, in order of increasing dash length.  We define $V_{\rm eff, \mu}(k)$ to be $V_{\rm eff}(\bfk)$ but averaged over a shell in $k$-space such that $V_{\rm eff, \mu}(k) \equiv 2 \, {\cal A} L \, P_{\rm F, \mu}^2/\delta P_{\rm F, \mu}^2$, where ${P}_{\rm F, \mu} \equiv {P}_F/(1+g\mu^2)^2$,
\begin{equation}
\delta {P}_{\rm F, \mu} =   \left(\, \int_0^1 d\mu \, (1+g\mu^2)^4 \, {\rm var}[\widehat{P}_{\rm F}(k, \mu)]^{-1} \right)^{-1/2},
\end{equation}
and ${\rm var}[\widehat{P}_{\rm F}(\bfk)] \equiv {\rm cov}[\widehat{P}_{\rm F}(\bfk), \widehat{P}_{\rm F}(\bfk)]$.  Here, $\delta {P}_{\rm F, \mu}$ has been defined so that $\delta {P}_{\rm F, \mu}/{\cal N}_k^{1/2}$ is the precision at which $P_{\rm F, \mu}$ can be measured in a $k$-space shell with ${\cal N}_k$ independent modes.   

The value of $V_{\rm eff, \mu}(k)/[{\cal A} \, L]$ is near unity when a survey is sample variance-limited.  For a BOSS-like quasar survey in which $\bar{n}_{\rm eff} = 10^{-3}~$Mpc$^{-2}$ at $2 <z <3$, this quantity reaches a maximum of $0.5$ at the smallest wavevectors and falls off rapidly at $k\gtrsim 0.1~$Mpc$^{-1}$.  Even $\bar{n}_{\rm eff} = 10^{-2}~$Mpc$^{-2}$ -- roughly the maximum density that can be achieve in a quasar survey -- is not sample variance-limited at $k\gtrsim 0.1~$Mpc$^{-1}$.  Figure \ref{fig:Veff} also emphasizes that $\bar{n}_{\rm eff} \approx 10^{-3}~$Mpc$^{-2}$ marks the critical number density at which a survey becomes sample variance-limited at $k\lesssim 0.1~$Mpc$^{-1}$, such that the gains from a $3$D analysis become large.

The plotted quantity in Figure \ref{fig:Veff} can also be related to the $S/N$ of quasar survey to ${P}_{\rm F, \mu}$ in a $k$-space shell of width $\Delta k$.  This shell contains ${\cal N}_{k} \equiv k^2 \Delta k \, {\cal A} \, L/(2 \pi^2)$  independent samples, so that the $S/N$ ratio at which $P_{\rm F, \mu}$ can be measured in this shell equals 
\begin{eqnarray}
P_{\rm F, \mu}/\sigma_{\rm F, \mu} &=& 60 \, \left(\frac{V_{\rm eff, \mu}/{\cal A} \, L}{10^{-2}}\right)^{1/2} \, \left( \frac{L} {1 {\rm \,Gpc}} \, \frac{{\cal A}}{10^3 {\rm \, deg}^2}\right)^{1/2} \nonumber \\
& & \times \left( \frac{1+z}{4} \right)^{1.2} \, \left(\frac{k^3 \, [\Delta k/k]}{10^{-3} \, {\rm Mpc^{-3}}}  \right)^{1/2},
\end{eqnarray}
where $\sigma_{\rm F, \mu} \equiv \delta {P}_{\rm F, \mu} (k) / {\cal N}_k^{1/2}$ and the redshift dependence is approximate.

\subsection{Estimating Correlations on Real Data}
\label{ss:realspace}

In practice, it will be easier to measure $3$D Ly$\alpha$ correlations in configuration space via the $3$D correlation function.  In configuration space, the Fourier-space weights proposed in Section \ref{ss:weights} become the line-of-sight convolution of $w_n(r)$ -- the Fourier Transform of $\tilde{w}_n(k_{\parallel})$ -- with the flux field of sightline $n$.  Conveniently, $w_n(r)$ will be relatively localized in real-space because $\tilde{w}_n(k_{\parallel}) \approx \tilde{w}_n(0)$ for $k_{\parallel} \lesssim 0.5~$Mpc$^{-1}$.  
However, for estimates of the correlation function, a simpler weighting for each sightline of $\tilde{w}_n(0)$ is likely sufficient.    In fact, this weighting is identical to the full weighting to the extent that $P_{\rm los}$ and the $P_{\rm N, n}$ are white.  We find that ${\rm var}[\widehat{P}_{\rm F}]$ is not significantly increased with these simpler weights.

Real data will be more complicated than the idealized case of uniform noise that we have considered thus far.  For example, there will be high-noise regions in the spectra owing to sky lines, and there may also be holes in the data where damped systems have been excised.  We have expressed our weights in terms of the noise and the Ly$\alpha$ forest power spectra.  However, the suggested weights are equivalent to simply weighting each sightline by $(1 +\sigma_{{\rm N}}^2/\sigma_{\rm los}^2)^{-1}$, where $\sigma_{\rm N}^2$ and $\sigma_{\rm los}^2$ are respectively the variance of the noise and the Ly$\alpha$ forest, smoothed on a large enough scale that $P_{\rm los}$ is white.  This scale corresponds to $\gtrsim 10~$Mpc, and we find that the exact choice of the smoothing scale weakly affects ${\rm var}[\widehat{P}_{\rm F}]$.  Such weights can more readily be applied to real data.



\section{Quadratic estimator}
\label{sec:QE}

 This section derives the minimum variance estimator that is quadratic in $\delta_{\rm F}$ and compares it to the estimator derived in the Section \ref{sec:sensitivity}.
 To proceed, we decompose the covariance of the flux overdensity at $k_{\parallel}$ into a component that depends on the parameter we aim to measure, $P_{\rm F}({\bfk})$, and one that does not.  Namely,
\begin{equation}
  \left\langle \tilde{\delta}_{{\rm F}, n} \, \tilde{\delta}_{{\rm F}, m} \right\rangle = C_{nm} +
     P_{\rm F}({\bfk}) \, X_{nm},
\end{equation}
where $\tilde{\delta}_{{\rm F}, n}$ is the Fourier transform of the flux along sightline $n$ and we consider a single $k_{\parallel}$,
\begin{eqnarray}
C_{nm}  &=& P(k_{\parallel}, \bfr_{nm}) + P_{{\rm N}, n} \, \delta_{n}^{m}, \\
X_{nm} &=& \left( \frac{\Delta k_\perp}{2\pi} \right)^2 \, \exp[i {\bfk}_\perp \cdot {\bfr}_{nm}] ,\\
P(k_{\parallel}, \bfr_{n m}) &\approx& \int \frac{d k_\perp}{2\pi} \, k_\perp \, P_{\rm F}(k_{\parallel}, k_{\perp}) \, J_0(k_\perp \, r_{nm}). \label{eqn:Pkr}
\end{eqnarray}
Here, $r_{nm} \equiv |\bfr_{nm}|$ is the transverse separation between sightlines $n$ and $m$, $\Delta k_\perp$ is the width of the Fourier space pixel in which we estimate $P_{\rm F}({\bfk})$, and equation (\ref{eqn:Pkr}) assumes that $P_{\rm F}({\bfk}) \, X_{nm} \ll P(k_{\parallel}, \bfr_{nm})$.

The minimum variance estimator for $P_{\rm F}({\bfk})$ that
is quadratic in $\tilde{\delta}_{{\rm F}, n}$ and unbiased is given by iterating
\begin{equation}
  \widehat{P}_{\rm F}^{\rm QE} =  [\widehat{P}_{\rm F}^{\rm QE}]_{\rm last} + \sum_{nm} \, Q_{n m} \, \tilde{\delta}_{{\rm F}, n} \tilde{\delta}_{{\rm F}, m} - {\rm tr} [\bfQ \, \bfC],\label{eqn:est}
\end{equation}
where $[\widehat{P}_{\rm F}^{\rm QE}]_{\rm last}$ is the previous iteration's estimate and
\begin{equation}
  \bfQ
   =  \left({\rm tr}
      \left[\bfC^{-1}\bfX \, \bfC^{-1} \bfX\right]\right)^{-1}
      \bfC^{-1} \bfX \, \bfC^{-1}.
      \label{eqn:Q}
\end{equation}

Equation (\ref{eqn:Q}) is derived by minimizing the variance of  $\widehat{P}_{\rm F}^{\rm QE}$ assuming $\delta_{\rm F}$ is Gaussian so that
\begin{equation}
 {\rm var}\left[\widehat{P}_{\rm F}^{\rm QE}  \right] 
  = 2 \, {\rm tr}\left[ \bfC \, \bfQ \,\bfC \,\bfQ\right],
  \label{eqn:varQE}
\end{equation}
and subject to the condition that it is unbiased so that ${\rm tr} [ \bfQ \, \bfX] = 1$.  Note that equation~(\ref{eqn:varQE}) does not include sample variance, which would contribute the term $2 \,P_{\rm F}^2$.

Our aim is to understand how the weights suggested in the previous section relate to the above minimum variance quadratic estimator.  To proceed, we set the off diagonal terms in $C_{nm}$ to zero (i.e., $P(k_{\parallel}, \bfr_{n m}) \rightarrow 0$ for $\bfr_{n m} \neq 0$).  The estimator is still unbiased with this approximation.  The motivation for this approximation is that the typical separation of quasars in such surveys is $r_\perp \gtrsim 10~$Mpc.  At these $r_\perp$ and at relevant $k_\parallel$, $P(k_{\parallel}, \bfr_\perp)$ is down by a factor of $\gtrsim 4$ from its $\bfr_\perp=0$ value.  In this ``zeroth-order'' approximation, $Q_{nm}$ becomes
\begin{equation}
Q_{nm}^{(0)}  = A \, \left(\frac{2\pi}{\Delta k_\perp}\right)^{2}  \;\frac{\exp[i  \,{\bfk}_\perp \cdot {\bfr}_{nm}]}{\left(P_{\rm los} +P_{{\rm N}, n} \right) \,  \left(P_{\rm los} +P_{{\rm N}, m} \right)},
\label{eqn:Qdiag}
\end{equation}
where $P_{\rm los} =  P(k_{\parallel}, 0)$ and 
\begin{eqnarray}
A  &=& \left( \sum_{n \,m} {\left(P_{\rm los} +P_{{\rm N}, n} \right)^{-1} \,  \left(P_{\rm los} +P_{{\rm N}, m} \right)^{-1}}  \, \right)^{-1},\\
&\approx&  \left(\sum_{n} \left(P_{\rm los} +P_{{\rm N}, n} \right)^{-1} \right)^{-2}.
\label{eqn:Qdiagnorm}
\end{eqnarray}
The last line assumes that the noise is uncorrelated between sightlines and neglects the terms with $m=n$, whose fractional contribution vanish as $N\rightarrow \infty$.
The final simplifications makes it apparent that this weighting is identical to that derived in Section \ref{ss:weights}, where we had implicitly made the same assumptions.  The estimator in Section \ref{ss:weights} is biased at the $N^{-1/2}$ level.  Subtracting the term ${\rm tr} [\bfQ^{(0)}  \bfC]$ from $\sum_{nm} \, Q_{n m} \, \tilde{\delta}_{{\rm F}, n} \tilde{\delta}_{{\rm F}, m}$ in equation (\ref{eqn:est}) corrects for this bias.    

Ignoring the ${\rm tr} [\bfQ^{(0)}  \bfC]$ correction, the corresponding estimator for measuring the correlation function $\xi(\bfr)$ follows from taking the Fourier transform of $Q_{nm}^{(0)}\, \tilde{\delta}_{{\rm F}, n} \tilde{\delta}_{{\rm F}, m}$ and, with the approximation in equation (\ref{eqn:Qdiagnorm}), is 
\begin{equation}
\widehat{\xi^{(0)}}(r) = \sum_{nm}^{{\rm fixed~}\bfr}\left[w_n(x) \star \delta_{\rm F}(\bfx_n) \right] \star  \left[w_m(x) \star \delta_{\rm F}(\bfx_m) \right],
\end{equation}
where $\star$ signifies a convolution along the line-of-sight.  This estimator is identical to that derived in a different manner in Section \ref{ss:realspace}.

\begin{figure}
{\epsfig{file=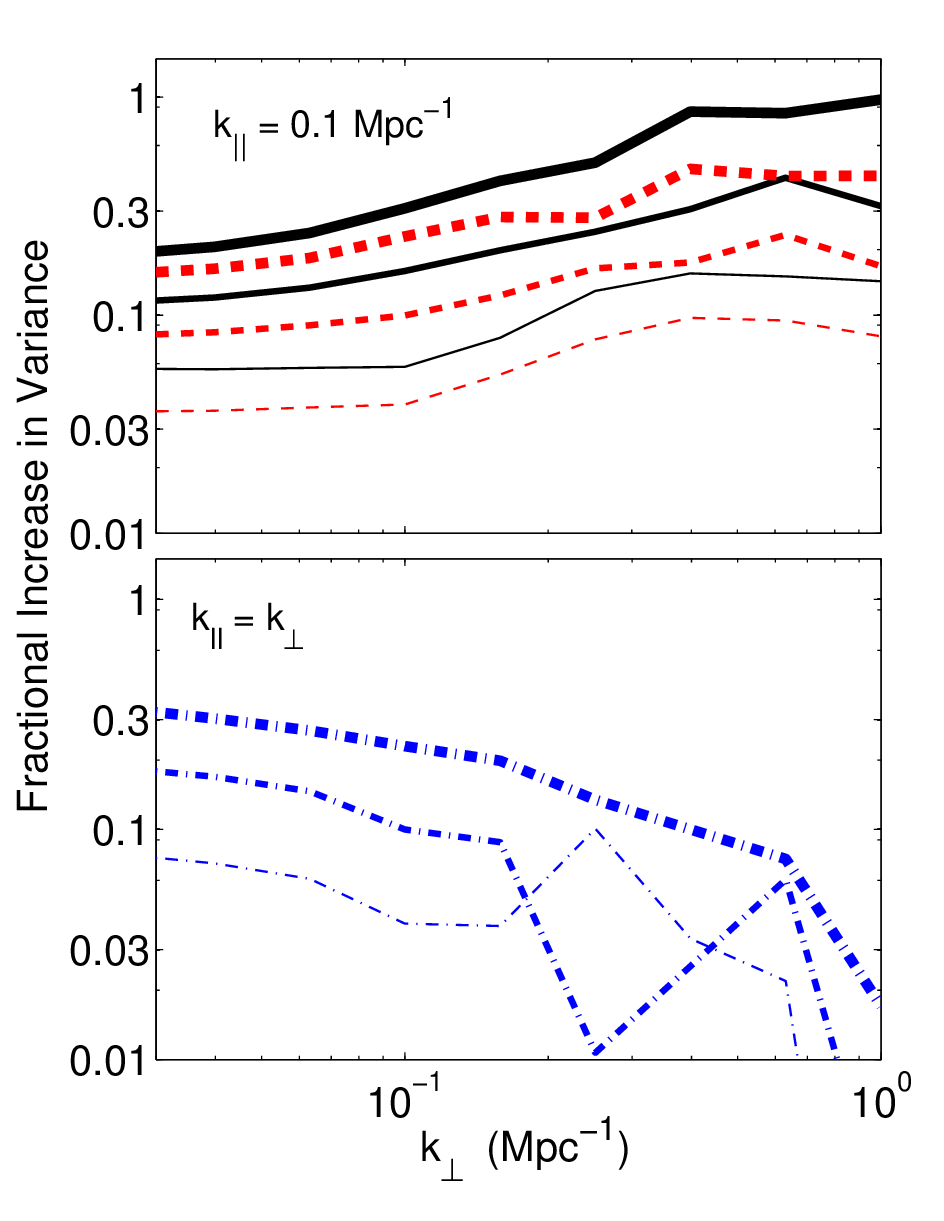, width=8.5cm}}
\caption{Top panel:  Fractional increase in variance for the $\widehat{P}_{\rm F}^{(0)}$ estimator relative to the minimum variance quadratic estimator, $\widehat{P}_{\rm F}^{\rm QE}$, as a function of $k_{\perp}$.   In order of increasing thickness, the curves are for $\bar{n} = 10^{-3}, \; 3 \times 10^{-3}$, and $10^{-2}~$Mpc$^{-2}$.  The black solid curves are the noiseless case, and the red dashed curves include noise as described in the text.  These calculations assume $k_{\parallel} = 0.1~$Mpc$^{-1}$, are performed in a $200 {\rm \; Mpc} \times200~$Mpc region, and do not include sample variance.  Bottom Panel:  The blue dot-dashed curves are the same as the red dashed curves in the top panel but with $k_\parallel = k_\perp$. \label{fig:QEcomp} } 
\end{figure}

Figure \ref{fig:QEcomp} plots the fractional difference between the variance of $\widehat{P}_F^{(0)}$ and that of the minimum variance quadratic estimator $\widehat{P}_F^{\rm QE}$, both calculated using equation (\ref{eqn:varQE}) but with the appropriate $\bfQ$.  The top panel fixes $k_{\parallel}$ at $0.1~$Mpc$^{-1}$ and varies $k_{\perp}$, and the bottom panel sets $k_{\parallel} = k_{\perp}$.  The solid curves in the top panel show the fractional difference in the zero-noise limit ($P_{{\rm N}, n} = 0$), which is the case where the fractional difference is largest.  In order of increasing thickness, the solid curves represent $\bar{n} = 10^{-3}, \; 3 \times 10^{-3}$ and $10^{-2}~$Mpc$^{-2}$.   At the smallest $k_\perp$ shown in the top panel, the estimator variance for $\widehat{P}_F^{(0)}$ is only $6\%$ larger for the case with $\bar n = 10^{-3}~$Mpc$^{-2}$ (thin solid curve).  This difference becomes $10\%$ and $20\%$ for $\bar n = 3 \times 10^{-3}~$Mpc$^{-2}$ and $10^{-2}~$Mpc$^{-2}$, respectively (thicker solid curves), and it increases with $k_\perp$.  These estimates for the fractional increase in variance do not include sample variance, which would further decrease the plotted fraction (especially for the highest $\bar n$ and smallest $k$).

The red dashed curves in the top panel of Figure \ref{fig:QEcomp} are the same as the black solid curves except that they include noise.  In particular, these curves assume a power-law luminosity function with slope $-2$, $z=2.5$, $P_{{\rm N}, n} = 1~$Mpc$^{-1}$ where quasar $n$ is at the limiting magnitude of the survey, and $S/N\propto {\rm ~flux}$.  These choices result in $\bar{n}_{\rm eff}$ being reduced by a factor of $0.7$ compared to the noiseless case, and the red dashed curves are suppressed relative to the black solid curves by a comparable factor.   We find a similar correspondence for other noise models. 

One can improve upon our crude diagonal approximation for $\bfC$ in an iterative manner.  In particular, maintaining only the diagonal elements in $\bfC$ when inverting can be thought of as the lowest order approximation for $\bfC^{-1}$ in equation~(\ref{eqn:Q}).  There are iterative methods that can be applied to achieve higher order corrections (such as the Jacobi method, Neumann iteration, or the Gauss-Seidel method).   We advocate for the Gauss-Seidel method here because it is guaranteed to converge since $\bfC$ is Hermitian, and we have indeed verified that it converges.  This method yields after the $i^{\rm th}$ iteration the estimate for the inverse of $\bfC$ given by
\begin{equation}
[\bfC_{\rm GS}^{(i)}]^{-1} = \boldsymbol{L}^{-1} \, \left(\boldsymbol{I} - \boldsymbol{U} \, [\boldsymbol{C}_{\rm GS}^{(i-1)}]^{-1} \right),
\end{equation}
where $\boldsymbol{L}$ and $\boldsymbol{U}$ are the lower and strictly upper diagonal components of $\bfC$.\footnote{The inversion of the lower diagonal matrix $\boldsymbol{L}$ requires at most $N^2$ operations.}  We initialize the iteration with $C^{(0)}_{{\rm GS}, nm} = C_{nm} \,\delta_{n}^{m}$ so that the zeroth order iteration yields $Q^{(0)}$ (eqn. \ref{eqn:Qdiag}) and the associated estimator, $\widehat{P}_{\rm F}^{(0)}$.  

The first order $\bfQ$ down-weights sightlines that fall near one another.  It is instructive to write the first order result using the simpler Jacobi iteration method.  In the Jacobi method, the first order approximation for the $\bfC^{-1}$ is 
\begin{equation}
[\bfC^{(1)}_{\rm J}]^{-1} =  [\bfC^{(0)}_{\rm J}]^{-1} (2\, \bfC^{(0)}_{\rm J} - \bfC)  \,  [\bfC^{(0)}_{\rm J}]^{-1}.  
\end{equation}
Being cavalier with the normalization, the next order weights in the Jacobi iteration scheme are given by
\begin{eqnarray}
Q_{{\rm J}, nm}^{(1)} 
& = & \tilde{w}_n \tilde{w}_m \, e^{i \bfk_\perp \cdot \bfr_{nm}} \left(1- \tilde{S}_n\right) \left(1 - \tilde{S}_m^* \right),   \label{eqn:QJ}
\end{eqnarray}
where  
\begin{eqnarray}
\tilde{S}_i(\bfk) &=& \sum_{\forall j, j\neq i} \frac{P(k_{\parallel}, \bfr_{ij})}{P_{\rm los}(k_{\parallel}) + P_{{\rm N}, j}}\exp[-i \bfk_\perp \cdot \bfr_{ij}].
\end{eqnarray}
The average of $S_i$ in a survey equals $P_{\rm F}(\bfk)/[\bar{n}_{\rm eff}^{-1} P_{\rm los}]$.  
If the number of quasars within $|\bfk_\perp \cdot \bfr_{ij}| \lesssim 1$ of quasar $i$ is larger than for other quasars, $S_i$ will also be larger, suppressing the weight given to this quasar.   However, if all quasars have a similar number of quasars within $|\bfk_\perp \cdot \bfr_{ij}| \lesssim 1$, then the $S_i$ would not vary strongly among the sightlines and this correction would be of minimal importance.

\section{Survey Considerations}
\label{sec:considerations}

\subsection{Galaxies}

\begin{figure}
\epsfig{file=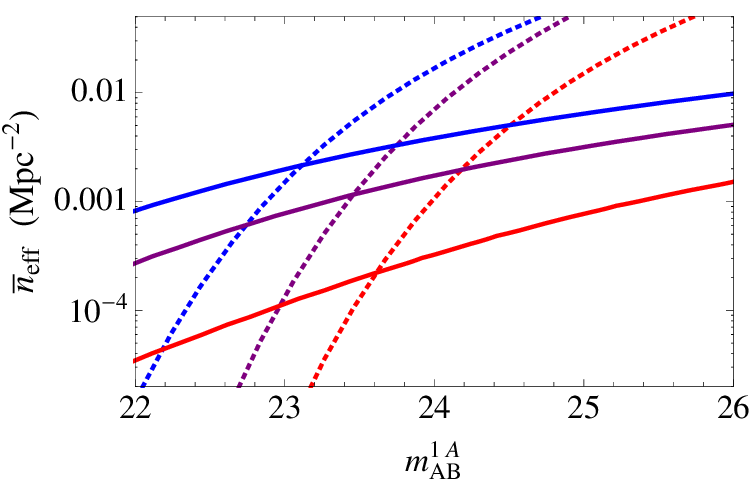, width=8cm}
\caption{$\bar{n}_{\rm eff}$  as a function of $m_{\rm AB}^{1A}$ for a Ly$\alpha$ forest survey that uses either galaxies or quasars.  These curves assume $m^{\rm lim}_{\rm AB} = m_{\rm AB}^{1A}$, where $m^{\rm lim}_{\rm AB}$ is the survey limiting magnitude. The solid curves show $\bar{n}_{\rm eff}$ for quasars at  $z=2, 3$, and $4$ (from top to bottom).  The dashed curves show this quantity for galaxies at $z=2, 3$, and $4$ (from left to right).  Here, $\bar{n}_{\rm eff}$ is calculated at $k_{\parallel} = 0.1~$Mpc$^{-1}$, but its $k_{\parallel}$ dependence is extremely weak.  \label{fig:Neffgal}}
\end{figure}

The apparent flatness of the faint end of the quasar luminosity function does not facilitate using observations of quasars with $L < L_*$ to obtain a denser sample of Ly$\alpha$ forest spectra.  However, galaxies can be used to supplement $\bar{n}_{\rm eff}$ in very deep surveys.  The dashed curves in Figure \ref{fig:Neffgal} represent $\bar{n}_{\rm eff}$ for a spectroscopic galaxy survey at $z=2, 3$, and $4$ (from left to right), assuming that $m^{\rm lim}_{\rm AB} = m_{\rm AB}^{1A}$.  These curves are calculated from the luminosity function of Lyman-break galaxies in \citet{2007ApJ...670..928B} for $z\geq3$, assuming that there is no evolution in the ultraviolet luminosity function between $z=2$ and $3$ \citep{reddy09}.  The galaxy magnitudes are for rest-frame $1600~$\AA.  The solid curves are the same but for a quasar survey at these redshifts.  Galaxies begin to aid the sensitivity when $m_{\rm AB}^{1A} \approx 23$ at $z=2$ and $m_{\rm AB}^{1A} \approx  23.5$ at $z=4$.  Including galaxies results in a dramatic increase in $\bar{n}_{\rm eff}$ once a survey pushes beyond these magnitude thresholds (Table \ref{table:neff}).  

The non-smooth continuum of galaxies' spectra will add an additional source of noise, but will not bias the estimate of $P_{\rm F}$ as long as correlations within a single sightline are neglected and the mean continuum can be removed (and even these mistakes can be isolated; Section \ref{sec:systematics}).  If the power in galaxies' continuum spectra is smaller than $\bar n/\bar n_{\rm eff} \times P_{\rm los}$, using galaxies in addition to quasars has the potential to significantly improve a survey's sensitivity.

\subsection{Cross Correlation}

It is also possible to cross-correlate Ly$\alpha$ forest skewers with some other tracer of large-scale structure.   This could be done using a galaxy survey or with the quasars in the Ly$\alpha$ survey themselves.  We express the overdensity field of these tracers as $\delta_g$ and their average $3$D number density as $\bar{n}_{\rm g, 3D}$.  The average cross power between these signals is $\widehat{P}_{\rm F g} (k) \equiv \langle \tilde{\delta} (k) \, \tilde{\delta}_{\rm g} (k) \rangle$ and the variance on an estimate of this signal is 
\begin{equation}
{\rm var}[\widehat{P}_{\rm F g}(\bfk)] =  P_{\rm F g}(\bfk)^2 + P_{\rm tot}(\bfk) \, \left(P_{\rm g}(\bfk) + \bar{n}_{\rm g, 3D}^{-1} \right),
\end{equation}
where $P_{\rm F g}$ is the cross-power spectrum between $\delta_{\rm F}$ and $\delta_{\rm g}$ and $P_{\rm g}$ is the auto-power of $\delta_{\rm g}$.  If we again assume the weight factorizes, weights given by equation (\ref{eqn:wn}) also maximize the $S/N$ in this case.  

There are three scenarios for which the cross power could be an interesting measurement: (1) the cross-correlation is more sensitive to $P^{\rm lin}_\delta$ than the Ly$\alpha$ forest auto-correlation, (2) it could be used to separate different contributions to the Ly$\alpha$ flux power or to measure the bias of the galaxies, or (3) it could be a systematic check for both the Ly$\alpha$ and the other survey.

Cross correlation is more sensitive than the Ly$\alpha$ forest auto-correlation when
\begin{equation}
n_{\rm g, 3D} > \bar{n}_{\rm eff} \, P_{\rm los}^{-1}(k_{\parallel}) \, b^2/b_g^2,
\label{eqn:cccond}
\end{equation}
where we have assumed that the shot noise and aliasing terms are the dominant sources of variance for both surveys.  For $\bar{n}_{\rm eff} = 10^{-3}~$Mpc$^{-2}$ (characteristic of BOSS), $b = 0.3$, and $P_{\rm los} = 0.5$ (where the latter two numbers are characteristic of the $z=3$ Ly$\alpha$ forest), condition (\ref{eqn:cccond}) becomes $n_{\rm g, 3D} > 2 \times 10^{-5}~$Mpc$^{-3}$ for a tracer with $b_g =3$.  In addition, at scales where the other survey is instead sample variance-limited, the $S/N$ in cross correlation will be at least as large as in the Ly$\alpha$ forest survey's auto-correlation.

Inequality (\ref{eqn:cccond}) requires a number density that is a factor of $\approx 10$ higher than the $3$D quasar density that BOSS aims to measure.  However, even though the $(S/N)^2$ in the cross correlation is then a factor of $\approx 10$ below the $(S/N)^2$ in the Ly$\alpha$ forest auto correlation, this should be still be sufficient to measure the quasar bias (and to much better precision than is possible with the quasar auto-power).   A more promising route may be to cross correlate with a separate high-redshift galaxy survey.  As an example, the Hobby-Eberly Telescope Dark Energy Experiment (HETDEX)\footnote{\url{http://www.as.utexas.edu/hetdex/}} aims to find spectroscopically $1~$million Ly$\alpha$ emitters between $1.8 < z < 3.8$ over $200~$deg$^2$, which works out to an average density of $\bar{n}_{\rm g, 3D} \approx 2\times 10^{-4}~$Mpc$^{-3}$ and easily satisfies criterion (\ref{eqn:cccond}) when correlated with a BOSS-like Ly$\alpha$ forest survey.  

If criterion (\ref{eqn:cccond}) is satisfied, the $S/N$ is greater than or equal for the galaxy survey's own auto-power compared to its cross power with the Ly$\alpha$ forest survey.  Thus, the cross power may be most interesting as a systematics check as well as to investigate the sources of fluctuations that contribute to $\delta_{\rm F}$.

\subsection{Survey Strategy}
\label{sec:strategy}

\begin{figure*}
\epsfig{file=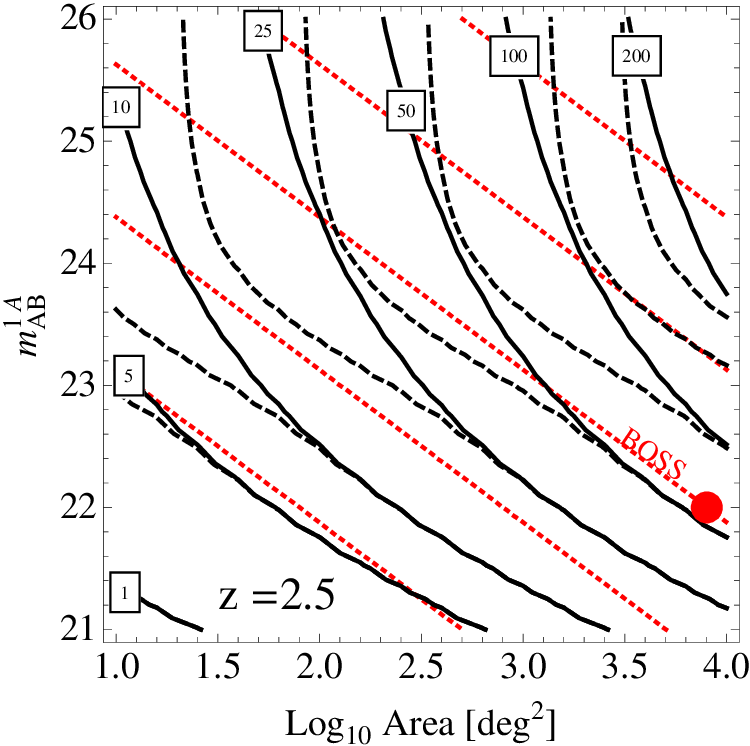, width=8cm}\epsfig{file=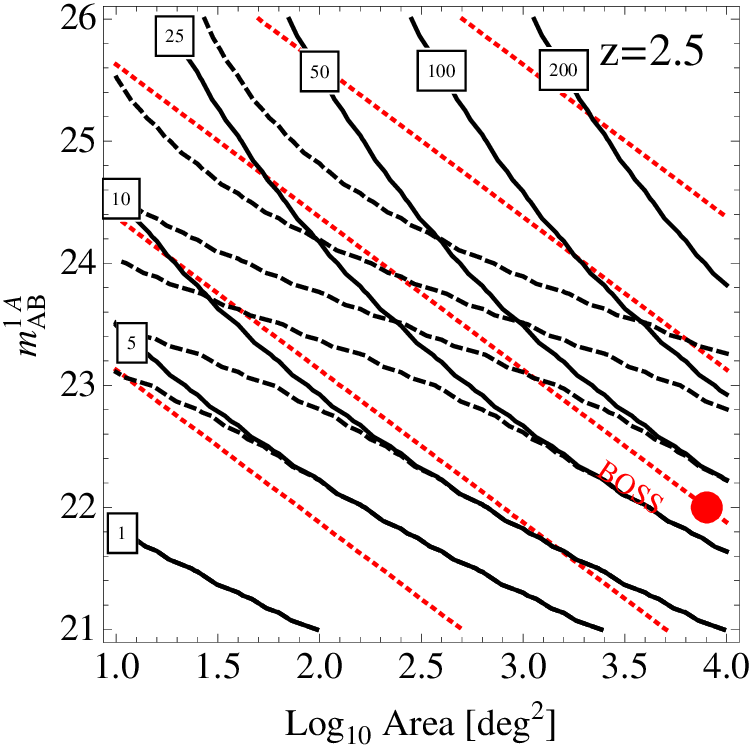, width=8cm}
\caption{Sensitivity to $P_{\rm F}$ at $k = 0.1~$Mpc$^{-1}$ (left panel) and $k = 0.5~$Mpc$^{-1}$ (right panel) as a function of survey area and limiting magnitude, where we have assumed that $m^{\rm lim}_{\rm AB} = m_{\rm AB}^{1A}$, $L \approx 500~$Mpc, and $z=2.5$.  The black solid curves are contours of constant sensitivity for a quasar survey and the often overlapping black long-dashed curves are at the same sensitivity contours as the black curves but for a survey that also includes galaxies down to the same magnitude.  The framed labels associated with each solid curve quote the $S/N$ on $P_{\rm F, \mu}$ in bins of width $\Delta k/k = 0.2$.  The red circle represents the approximate specifications of the BOSS quasar survey.  The red short-dashed curves are contours of constant ${\cal M} \equiv {\rm Survey~Area}/({\rm limiting ~flux})^2$.  For fixed duration of the observing campaign on an instrument, ${\cal M}$ does not depend on the survey strategy.  The red short-dashed curves correspond to $10^{-2}$, $10^{-1}$, $1$, $10$, and $100$ times the ${\cal M}$ of the BOSS survey.   \label{fig:senscont}}
\end{figure*}

Figure \ref{fig:senscont} illustrates the survey strategy tradeoffs in the sky coverage versus magnitude-limit plane for $z=2.5$.  The black solid curves in both panels are contours of constant sensitivity to $P_{\rm F}$ for a quasar survey.  The black long-dashed curves are for a survey with quasars plus galaxies and at the same sensitivity as the black solid curves.  The framed labels associated with each solid curve show the $S/N$ of a measurement of $P_{\rm F, \mu}$ in a $k$-space shell of width $\Delta k/k = 0.2~$.  This shell is centered on $k = 0.1~$Mpc$^{-1}$ (left panel) or on $k = 0.5~$Mpc$^{-1}$ (right panel).  

The red short-dashed curves are contours of constant ${\cal M} \equiv {\rm Survey~Area}/({\rm limiting ~flux})^2$.  For fixed duration of the observing campaign on an instrument, ${\cal M}$ is a constant that does not depend on the survey strategy (assuming that the limiting flux decreases as exposure time on a field to the $1/2$ power).  The curves in Figure \ref{fig:senscont} correspond to ${\cal M}$ of $10^{-2}$, $10^{-1}$, $1$, $10$, and $100$ times ${\cal M}$ for the BOSS survey, whose approximate specifications are represented by the filled circle ($m^{\rm lim}_{\rm AB} = m^{1A}_{\rm AB} = 22$, ${\cal A} = 8000~$deg$^{2}$).  For the present specifications of BigBOSS, ${\cal M}$ is $\approx 7$ times larger than for BOSS.  \citet{mcdonald05b} suggested an extension of the WFMOS spectroscopic galaxy survey to include quasars \citep{glazebrook05} and would reach $m^{\rm lim}_{\rm AB} \approx24.5$.  WFMOS is a proposed $300~$deg$^2$ survey on the $8$~m Subaru telescope.  This hypothetical quasar survey would be only moderately more sensitive than BOSS at the considered scales.

The left panel in Figure \ref{fig:senscont} argues that the survey strategy for BOSS is close to optimal in the sense of minimizing ${\rm var}[\widehat{P}_{\rm F}]$ at $k = 0.1~$Mpc$^{-1}$:  The sensitivity contours have a similar shape to the red short-dashed curve that intersects the BOSS point.  Similar conclusions hold for the sensitivity at $k=0.5~$Mpc$^{-1}$ (right panel, Fig. \ref{fig:senscont}).  In general, the optimal strategy for a quasar survey corresponds to integrating deep enough on each field (and having enough fibers/slits) to reach quasars with luminosities of $\approx L_*$, where $L_*$ is the characteristic luminosity in the quasar luminosity function.  While Figure \ref{fig:senscont} assumes that $m_{\rm AB}^{\rm lim} = m_{\rm AB}^{\rm 1A}$, the vertical axis can also be thought of as a function of $\bar{n}_{\rm eff}$, noting that $m_{\rm AB}^{\rm 1A} = m_{\rm AB}^{\rm lim} = \{22,~ 24,~ 26\}$ corresponds to $\bar{n}_{\rm eff} = \{0.5, ~3, ~7 \} \,  {\rm ~Mpc}^{-2}$ at $z=2.5$ for a survey that only includes quasars (Table \ref{table:neff}).

At the considered wavevectors and fixed ${\cal M}$, the sensitivity to $P_{\rm F}$ depends little on a survey's depth, at least for $m_{\rm AB}^{\rm lim}  < 23$.  This result owes to the shape of the bright end of the luminosity function.  However, we favor the survey strategy that is deep enough to reach down to $L_*$ because deeper quasar surveys will be able to better handle systematics (because they have higher $S/N$ per mode) and they rely less on the noise decreasing as the square root of the number of modes (which can be invalidated by non-Gaussian effects; \citealt{meiksin99}).  Technically, a shallower but wider survey strategy will be more sensitive at smaller $k$ than shown here because it samples more modes. 

The sensitivity gains of a deep survey that uses the spectra from galaxies in addition to quasars are not always significant.  Compare the long-dashed and solid curves in Figure \ref{fig:senscont}.  Including galaxies tends to improve the statistical sensitivity of surveys for limiting magnitudes greater than $23-24$, and the denser sample of sightlines that galaxies provide is more helpful at $k=0.5~$Mpc$^{-1}$ compared to $k=0.1~$Mpc$^{-1}$ (Fig. \ref{fig:senscont}).  However, the sensitivity is not always significantly improved over a quasar survey covering a larger fraction of the sky (fixing ${\cal M}$).  The gains in sensitivity from including galaxies may not always be sufficient to outweigh the added difficulty of removing the galaxies' more complex continua.  

\begin{figure}
\epsfig{file=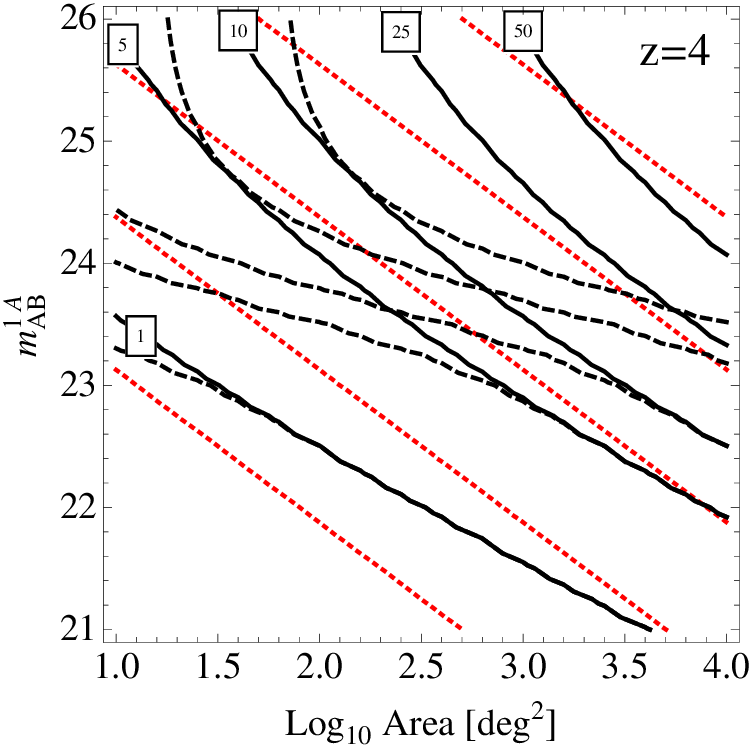, width=8cm}
\caption{Same as Figure \ref{fig:senscont} but at $z=4$ and for $k = 0.1~$Mpc$^{-1}$.\label{fig:senscontz4}}
\end{figure}

It is much more difficult to measure $P_{\rm F}$ at $z=4$.  However, some of the science drivers of a measurement at $z=4$ may not require a precision measurement, and even an order unity measurement at this redshift would be interesting \citep{mcquinn10}.  Figure \ref{fig:senscontz4} quantifies the prospects for a measurement of $P_{\rm F}$ at $z=4$.  A spectroscopic quasar survey reaching $m_{\rm AB}^{1A} = 22$ (the same depth as BOSS) would be able to achieve $S/N \approx 3$ at $k=0.1~$Mpc$^{-1}$ in shells of width $\Delta k/k = 0.2$.  A survey reaching $m_{\rm AB}^{1A} = 23$ would achieve $S/N \approx 10$.  

Thus far we have not discussed how to select quasars as a function of redshift:  A survey with a multi-object spectrograph would in practice have to decide whether to obtain the spectra of a quasar at $z=2$ or $z=3$, and this choice would affect its sensitivity.  In the aliasing noise-limited regime, the total $[S/N]^2$ on a scale is $ \propto \int dz \, \bar{n}_{\rm eff}^2$.  Because this integral is quadratic in $\bar{n}_{\rm eff}$, maximizing $\bar{n}_{\rm eff}$ over a limited interval in redshift also maximizes the $S/N$.  However, once sample variance limits a survey at the scales of interest, it makes sense to broaden the redshift width of the quasar selection function.  In practice, a small fraction of slits or fibers per degree compared to the total number used in modern spectroscopic galaxy surveys is required to select all $z>2$ quasars down to a survey's limiting flux ($\approx 10-100$).   Therefore, the decision of which quasars to target may often be moot.

\subsection{Cosmological and Astrophysical Constraints}
\label{ss:cosmo}

Ultimately, one wants to use $3$D Ly$\alpha$ forest measurements to constrain cosmological parameters, ionizing background models, and the reionization history.  This section briefly discusses how well these quantities can be constrained.  To do so, we calculate the Fisher matrix defined as (e.g., \citealt{tegmark97})
\begin{equation}
F_{ij} \equiv - \left \langle \frac{\partial^2 \ln{\cal L}}{\partial \lambda_i \partial \lambda_j} \right \rangle = \frac{1}{2} \sum_{\forall \bfk} P_{\rm tot}(\bfk)^{-2}  \, \frac{\partial P_{\rm F}(\bfk)}{\partial \lambda_i}  \frac{\partial P_{\rm F}(\bfk)}{\partial \lambda_j},
\label{eqn:fish}
\end{equation}
where ${\cal L}$ is the likelihood of the model given the data, the second equality assumes Gaussianity, and the $\lambda_i$ are the parameters we want to constrain.  The summation is over all independent modes.  Given this parameter set, the forecasted $1 \,\sigma$ uncertainty on the parameter $\lambda_i$ is $[\matF^{-1}]_{ii}^{1/2}$.

We perform our calculations for a survey at $z=2.5$ and with a base parameter set that is given by the amplitude of $P_{\rm F}$, its tilt and running with pivot at $k = 0.1~$Mpc$^{-1}$, the redshift-space distortion parameter $g$, the angular diameter distance $D_A$, and the Hubble expansion parameter $H(z)$.  The former four parameters we treat as nuisance parameters.  Finally, we omit the first three line-of-sight modes since these are likely to be contaminated by continuum and mean flux errors, as motivated in \citet{mcdonald05}.  The results depend weakly on this assumption.

Some of the constraint from our base parameter set on $D_A(z)$ and $H(z)$ owes not just to the baryon acoustic oscillation (BAO) features, but also to Alcock-Paczynski type effects.  Since it might be the case that the continuum under the BAO is contaminated by other  effects, one may not want to use information that derives from this broad-band power.  To guarantee that the constraint owes to the BAO features, we also provide a more conservative estimate in which we subtract from $P_{\rm F}$ a flux power spectrum that does not include the BAO prior to calculating $\partial P_{\rm F}/\partial \lambda_i$.  In what follows, we quote both constraints.

\begin{figure}
\rotatebox{-90}{\epsfig{file=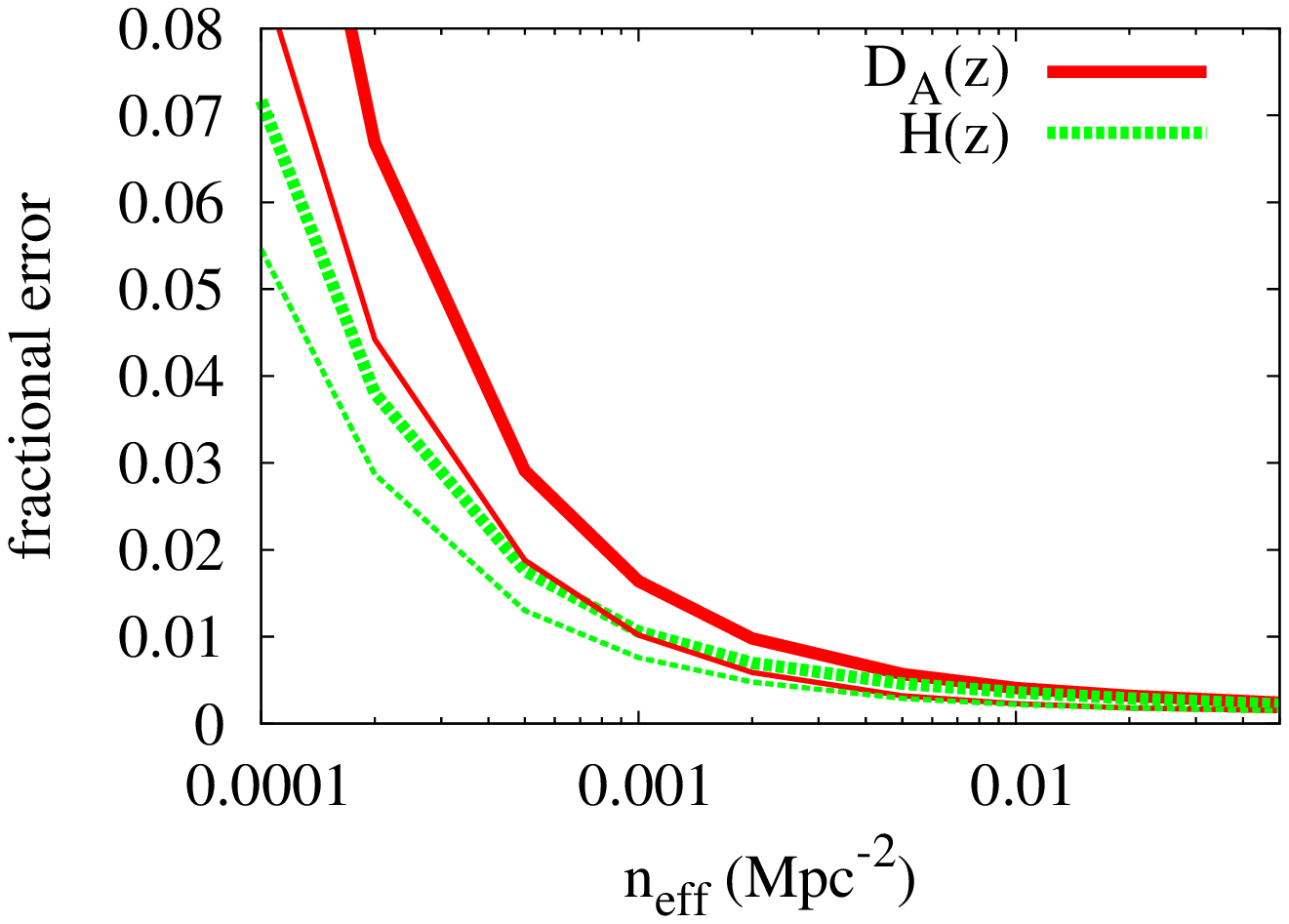, height=8.8cm}}
\caption{Fractional error on the angular diameter distance and Hubble expansion parameter at $z=2.5$ as a function of $\bar n_{\rm eff}$.  The curves are calculated for a survey with $L = 1~$Gpc and ${\cal A} = 10^4~$deg$^2$, but their amplitude scales as $(L \, {\cal A})^{-1/2}$.  The thick curves represent a conservative estimate, and the thin curves represent an optimistic estimate, as described in the text. \label{fig:cosmo_const}}
\end{figure}

Figure \ref{fig:cosmo_const} plots the constraints on $D_A(z)$ and $H(z)$ at $z=2.5$ as a function of $\bar{n}_{\rm eff}$.  The curves are for a survey with $L = 1~$Gpc and ${\cal A} = 10^4~$deg$^2$.  The thin curves represent the optimistic predictions using the fiducial estimate for $\matF$, and the thick curves are from the more conservative estimate that subtracts the continuum underneath the BAO.  A BOSS-like survey with $A \approx 10^4~$deg$^2$, $L\approx 1~$Gpc, and $\bar{n}_{\rm eff} \approx 5\times 10^{-4}~$Mpc$^{-2}$ ($m_{\rm AB}^{\rm lim} =22$) can constrain $D_A$ at $z=2.5$ to fractional precision of $\approx 0.02- 0.03$, and one like BigBOSS where $\bar{n}_{\rm eff} \approx 1.5\times 10^{-3}~$Mpc$^{-2}$ ($m_{\rm AB}^{\rm lim} =23$) to a fractional precision of $\approx 0.010-0.015$.\footnote{The angular diameter distance is generally better constrained in surveys of large-scale structure for geometric reasons.  However, the strength of redshift-space distortions in Lyman-$\alpha$ forest surveys enhances the line-of-sight power, boosting the sensitivity to $H(z)$.}  The constraints we found on $H(z)$ and $D_A(z)$ are comparable to those in \citet{mcdonald05}.  Figure \ref{fig:cosmo_const} allows one to estimate the sensitivity of a survey of arbitrary volume since the error scales as $(L \, {\cal A})^{-1/2}$.

The precision at which $D_A(z)$ is measured from the Ly$\alpha$ forest provides a constraint on cosmological parameters, such as the curvature space density, $\Omega_k$.  In particular, a measurement of $D_A(z)$ at $z\approx 2.5$ differenced with its value at $z=1100$ -- which is tightly constrained by cosmic microwave background measurements -- provides a robust measure of $\Omega_k$ (at least under the assumption that dark energy is negligible at $z>2$).   If the principle uncertainty is in the angular diameter distance to the Ly$\alpha$ forest, as is likely to be the case, this translates to the constraint $\delta \Omega_k \approx 0.26 \, \delta D_A/D_A$ at $z=2.5$.  Thus, a measurement of $\Omega_K$ to the precision $\approx 3 \times 10^{-3}$ is possible with a BigBOSS-like survey, an improvement over the present-day limit of $|\Omega_K| < 1\%$ that assumes $\Lambda$CDM \citep{komatsu10}.  

Ionizing background and temperature fluctuations can also be measured in a $3$D Ly$\alpha$ forest analysis.  To gauge how well these contributions can be constrained, in addition to our base set of parameters, we also include a scale-invariant contribution to $P_{\rm F}$ that scales as $k^{-3}$, a component that scales as $k^{-2}$, and a white component that scales as $k^0$ (all spherically symmetric).  The contribution that scales as $k^{-2}$ is motivated by the expected scaling for intensity fluctuations at wavevectors larger than one over the mean free path of $1~$Ry photons (essentially all accessible $k$ at $z=2.5$; \citealt{mcquinn10}), and the scale-invariant form approximates some of the temperature models in \citet{mcquinn09}.  We have set the normalization of each component to be such that it is the same as in our simple model for $P_{\rm F}$ at $(k_\parallel, k_\perp) = (0.1, 0)~$Mpc$^{-1}$.   This normalization is comparable to that of physically motivated models for temperature and intensity fluctuations presented in \citet{mcquinn10}.  For a BOSS-like survey, the normalization of the $k^{-2}$ component is constrained to $10\%$, the normalization of the $k^{-3}$ component to $3\%$, and the normalization of the white component to $0.03\%$.  We have also examined the constraints for a shallower survey with $\bar n_{\rm eff} = 10^{-4}~$Mpc$^{-2}$, $L=1~$Gpc, and ${\cal A}=10^4~$deg$^2$ (as could be obtained by BigBOSS if it targets $z\approx 4$).  This case yields $40\%$ and $12\%$ constraints on the $k^{-2}$ and $k^{-3}$ components, respectively. 

\section{Systematic Concerns}
\label{sec:systematics}

This section briefly discusses systematics that are known to be serious for line-of-sight Ly$\alpha$ forest analyses in the context of $3$D analyses.  We focus on continuum fitting errors, mistakes in the mean flux estimate, and the damping wings of high column-density absorption systems. A pervading theme of this discussion is the advantages of a $3$D analysis in dealing with these systematics.

Measurements of $P_{\rm F}(\bfk)$ must remove the structure in the quasar continuum in order to achieve an unbiased measurement of $\delta_{\rm F}$.  However, continuum fluctuations do not have to be removed perfectly since they only act as a source of noise as long as the error in subtracting them does not correlate between different sightlines \citep{2002MNRAS.329..848V}.  (In addition, any systematic error introduced in this removal can be excised at a later step as described below.)  The extra continuum power that does not correlated between the different continua, $P_{\rm los}^{\rm cont}$, would contribute an extra term in $P_{\rm tot}$ equal to $\bar n^{-1} P_{\rm los}^{\rm cont}$, where again ${\rm var}[\widehat{P}_{\rm F}] = 2  \,P_{\rm tot}^2$, assuming that line-of-sight correlations are omitted.  At $k_{\parallel} > 0.1$~Mpc$^{-1}$, \citet{mcdonald05b} found that the mean quasar continuum had power equal to $<1\%$ of $P_{\rm F}$.   Thus, continuum fluctuations will not significantly increase the variance (by increasing the amount of line-of-sight power) at $k_\parallel > 0.1$~Mpc$^{-1}$, but could be of more importance at smaller wavevectors.  

The redshift evolution of the mean flux $\langle F \rangle$ -- which can be quite substantial over $L \sim 500~$Mpc -- can also contribute spurious power at small $k_{\parallel}$.  Accurately estimating $\langle F \rangle$ is known to be crucial for interpreting line-of-sight measurements.  It also has the potential to be an even more important issue for $3$D measurements, where the size of fluctuations one aims to measure are significantly smaller.  For example, the dimensionless $3$D power at $k = 0.1~$Mpc$^{-1}$ is $\Delta_{\rm F}^2 \sim 10^{-4}$, which would require that the mean flux evolution be known to precision $\ll 10^{-2}$ in order to not bias the measurement of $P_{\rm F}$ averaged over a shell centered at this $k$.  (Although, the power from the mean flux error would primarily impact purely line-of-sight modes.)  Traditionally, $\langle F \rangle$ is estimated by averaging the transmission at redshift $z$ from all the sightlines in a survey.  Given $N$ sightlines and if we want to estimate $\langle F \rangle$ in a region of size $\Delta \chi$, then this quantity can be estimated to precision
\begin{equation}
\sigma_{\langle F \rangle}^2 \approx \frac{(P_{\rm los} + P_{\rm N})} {\Delta \chi \; N}.
\end{equation}
Noting that the numerator on the right hand side is $\mathcal{O}(1)$, it takes $1000$ Ly$\alpha$ forest spectra to constrain $\sigma_{\langle F \rangle}$ in a $10~$Mpc region to reach the threshold of $10^{-2}$, which is much less than the total number of Ly$\alpha$ forest spectra in BOSS or BigBOSS.

However, stacking to obtain $\langle F \rangle$ requires that no mistakes are made.  For example, a systematic error in the continuum subtraction can lead to a biased estimate of $\langle F \rangle$.  Fortunately, in $3$D analyses, one has the ability to discern these effects at a later step in the analysis.  A mean flux error will principally lead to spurious power that only affects purely line-of-sight modes.  Thus, one can simply throw away the first $X$ modes along the origin in the analysis to remove these errors.   In addition, any bleeding to other modes owing to the complicated survey window may not be so worrisome because the error in the mean flux is at the percent-level such that the total variance contributed by this error is comparable to $\Delta_F^2$.  For example, the power in this bleeding for the $P_{\rm F}$ estimator described in Section \ref{sec:sensitivity} is suppressed by $N^{-1/2}$ relative to the mean flux-error power in the line-of-sight modes.   Errors in the mean flux also enter in convolution with $P_{\rm F}$, but this effect is less of a contaminant again given that the sizes of these errors are percent-level over $x \sim100~$Mpc.  For such an error, a percent of $\delta_{\rm F}$ is smoothed over $\Delta k_\parallel \sim x^{-1}$.  

Another systematic is the damping wing absorption from Lyman-limit and damped Lyman-$\alpha$ systems.  This absorption is generally not included in simulations of the Ly$\alpha$ forest, and it arises from dense, self-shielding systems that are not captured properly in almost all cosmological simulations anyway \citep{katz96, mcquinn11}.  These systems could alter the line-of-sight power at the $10\%$ level at $k_\parallel \approx 0.1~$Mpc$^{-1}$ \citep{mcdonald05}, and at the $100\%$ level at $k_\parallel = 0.01~$Mpc$^{-1}$ (Appendix \ref{ap:DLAs}).

 Appendix \ref{ap:DLAs} discusses a simple model for the power in these systems that qualitatively reproduces the numerical estimates for the effect of damping wings in \citet{mcdonald05} and that allows us to estimate their impact for $3$D analyses.  We show that much of the line-of-sight power from damping wings arises from the uncorrelated (shot) component of their power.  The shot contribution to the $3$D flux power from damping wings is much smaller than in the line-of-sight flux power spectrum, and, thus, the role of damping wings as a contaminant is reduced (Appendix \ref{ap:DLAs}).  However, we show that they still are likely to add power to $P_{\rm F}$ at the tens of percent level.  


\section{Conclusions}
\label{sec:discussion}

This paper studied issues relevant to upcoming $3$D Ly$\alpha$ forest surveys.  
We derived a simple formula for how to optimally weight sightlines with varying $S/N$ levels.  We found that simply weighting sightline $i$ by $(1 +\sigma_{{\rm N}, i}^2/\sigma_{\rm los}^2)^{-1}$, where $\sigma_{{\rm N}, i}^2$ and $\sigma_{\rm los}^2$ are respectively the variance of the noise and of the Ly$\alpha$ forest normalized flux in a $10~$Mpc region or greater, performs nearly as well as the weights specified by the minimum variance quadratic estimator.  These weights should be simple to apply to data even in the presence of real-world complications.  We derived an expansion (which involves only matrix multiplications) that converges to the minimum variance quadratic estimator and for which our suggested weights are the lowest order contribution.  We showed that the next term in this expansion has the intuitive behavior of suppressing the contribution from quasars that have an overabundance within $r_{\perp} \lesssim k_\perp^{-1}$.

We showed that the sensitivity of a spectroscopic survey to the Ly$\alpha$ forest flux power spectrum can be quantified by just a single number, $\bar{n}_{\rm eff}$ -- a noise-weighted number density of sources on the sky -- so that ${\rm var}[\widehat{P}_{\rm F}(\bfk)] = 2 \, (P_{\rm F} + \bar{n}_{\rm eff}^{-1} \,P_{\rm los})^2$.  While this number technically depends on the $k_\parallel$ of the wavevector in question, in practice this dependence is  extremely weak at $k_\parallel < 0.5~$Mpc$^{-1}$ because $P_{\rm los}$ is almost constant over these wavevectors.  These are the same wavevectors at which $3$D Ly$\alpha$ surveys have the potential to derive competitive constraints on cosmological parameters \citep{mcdonald07} and that are the most interesting for studying astrophysical sources of fluctuations in the Ly$\alpha$ forest \citep{mcquinn10}. 

We calculated $\bar{n}_{\rm eff}$ as a function of survey specifications for both quasar and galaxy surveys at different redshifts.  
For quasar surveys, it is difficult to achieve significantly higher $\bar{n}_{\rm eff}$ than $\approx 3 \times 10^{-3}~$Mpc$^{-2}$ (or $\approx 30$ per deg$^{2}$) at any redshift owing to the shallowness of the faint end of the quasar luminosity function.  A survey with $\bar{n}_{\rm eff} = 3\times10^{-3}~$Mpc$^{-2}$ is aliasing noise-limited at $k > 0.1~$Mpc$^{-1}$.  In this limit, the $S/N$ on $P_{\rm F}$ scales linearly with $\bar{n}_{\rm eff}$.

This paper also discussed survey strategy tradeoffs.  The previous results allowed for a simpler characterization of these tradeoffs than in prior studies.  We showed that a survey's sensitivity to the flux correlation function is maximized with the strategy of integrating on each field just long enough to achieve $S/N\approx 2$ in a $1\,$\AA\ pixel for an $L_*$ quasar.  However, we found that a shallower strategy but covering a wider field formally obtains a comparable sensitivity to $P_{\rm F}$ at $0.1 < k < 0.5~$Mpc$^{-1}$ (but with lower $S/N$ per mode), and it could even be more sensitive at smaller $k$.  While such low $S/N$ values may cause worry regarding the efficacy of continuum subtraction, we further argued that continuum removal errors as well as other systematics are likely to be less problematic in $3$D correlation analyses compared to in $1$D.  Lastly, we found that using the spectra from galaxies results in the statistical sensitivity being improved for surveys that achieve limiting magnitudes of $m_{\rm AB}^{1A} > 23-24$.  However, the improvements are not always significant when compared with a shallower but wider survey that only incorporates quasars.
 
We quantified as a function of $\bar n_{\rm eff}$ a survey's sensitivity to the angular diameter distance at the mean redshift of the survey, to the Hubble expansion rate, and to the sources of fluctuations that do not trace density.  Upcoming surveys have the potential to measure $D_A(z)$ and $H(z)$ to $1-2\%$, which would substantially improve constraints on the curvature of space-time and early dark energy models.  These surveys also have the potential to detect other contributions to the flux power at the percent-level and, thereby, constrain the level of temperature fluctuations relic from reionization processes and the properties of the extragalactic ionizing background.\\
 
 
We thank Andreu Font-Ribera and An\v{z}e Slosar for helpful comments on this manuscript, and Shirley Ho, Adam Lidz, Nic Ross, Jennifer Yeh, and Matias Zaldarriaga for useful discussions.  MM is supported by the NASA Einstein Fellowship.

\bibliographystyle{mn2e}
\bibliography{References}

\begin{thebibliography}{}

\bibitem[\protect\citeauthoryear{{Becker}, {Bolton}, {Haehnelt} \&
  {Sargent}}{{Becker} et~al.}{2011}]{becker10}
{Becker} G.~D.,  {Bolton} J.~S.,  {Haehnelt} M.~G.,    {Sargent} W.~L.~W.,
  2011, \mnras, 410, 1096

\bibitem[\protect\citeauthoryear{{Becker} et~al.,}{{Becker}
  et~al.}{2001}]{2001AJ....122.2850B}
{Becker} R.~H.,  et~al., 2001, \aj, 122, 2850

\bibitem[\protect\citeauthoryear{{Bouwens}, {Illingworth}, {Franx} \&
  {Ford}}{{Bouwens} et~al.}{2007}]{2007ApJ...670..928B}
{Bouwens} R.~J.,  {Illingworth} G.~D.,  {Franx} M.,    {Ford} H.,  2007, \apj,
  670, 928

\bibitem[\protect\citeauthoryear{{Cooray} \& {Sheth}}{{Cooray} \&
  {Sheth}}{2002}]{cooray02}
{Cooray} A.,  {Sheth} R.,  2002, \physrep, 372, 1

\bibitem[\protect\citeauthoryear{{Croft}, {Weinberg}, {Katz} \&
  {Hernquist}}{{Croft} et~al.}{1998}]{1998ApJ...495...44C}
{Croft} R.~A.~C.,  {Weinberg} D.~H.,  {Katz} N.,    {Hernquist} L.,  1998,
  \apj, 495, 44

\bibitem[\protect\citeauthoryear{{Fan} et~al.,}{{Fan}  et~al.}{2006}]{fan06}
{Fan} X.,  et~al., 2006, \aj, 132, 117

\bibitem[\protect\citeauthoryear{{Feldman}, {Kaiser} \& {Peacock}}{{Feldman}
  et~al.}{1994}]{feldman94}
{Feldman} H.~A.,  {Kaiser} N.,    {Peacock} J.~A.,  1994, \apj, 426, 23

\bibitem[\protect\citeauthoryear{{Glazebrook}, {Eisenstein}, {Dey}, {Nichol} \&
  {The WFMOS Feasibility Study Dark Energy Team}}{{Glazebrook}
  et~al.}{2005}]{glazebrook05}
{Glazebrook} K.,  {Eisenstein} D.,  {Dey} A.,  {Nichol} B.,    {The WFMOS
  Feasibility Study Dark Energy Team} 2005, astro-ph/0507457

\bibitem[\protect\citeauthoryear{{Hamilton}, {Rimes} \&
  {Scoccimarro}}{{Hamilton} et~al.}{2006}]{HRS06}
{Hamilton} A.~J.~S.,  {Rimes} C.~D.,    {Scoccimarro} R.,  2006, \mnras, 371,
  1188

\bibitem[\protect\citeauthoryear{{Hennawi} \& {Prochaska}}{{Hennawi} \&
  {Prochaska}}{2007}]{hennawi07}
{Hennawi} J.~F.,  {Prochaska} J.~X.,  2007, \apj, 655, 735

\bibitem[\protect\citeauthoryear{{Hopkins}, {Hernquist}, {Cox}, {Di Matteo},
  {Robertson} \& {Springel}}{{Hopkins} et~al.}{2006}]{hopkins06}
{Hopkins} P.~F.,  {Hernquist} L.,  {Cox} T.~J.,  {Di Matteo} T.,  {Robertson}
  B.,    {Springel} V.,  2006, \apjs, 163, 1

\bibitem[\protect\citeauthoryear{{Katz}, {Weinberg}, {Hernquist} \&
  {Miralda-Escude}}{{Katz} et~al.}{1996}]{katz96}
{Katz} N.,  {Weinberg} D.~H.,  {Hernquist} L.,    {Miralda-Escude} J.,  1996,
  \apjl, 457, L57+

\bibitem[\protect\citeauthoryear{{Komatsu} et~al.,}{{Komatsu}
  et~al.}{2011}]{komatsu10}
{Komatsu} E.,  et~al., 2011, \apjs, 192, 18

\bibitem[\protect\citeauthoryear{{Lidz}, {Faucher-Gigu{\`e}re}, {Dall'Aglio},
  {McQuinn}, {Fechner}, {Zaldarriaga}, {Hernquist} \& {Dutta}}{{Lidz}
  et~al.}{2010}]{lidz09}
{Lidz} A.,  {Faucher-Gigu{\`e}re} C.,  {Dall'Aglio} A.,  {McQuinn} M.,
  {Fechner} C.,  {Zaldarriaga} M.,  {Hernquist} L.,    {Dutta} S.,  2010, \apj,
  718, 199

\bibitem[\protect\citeauthoryear{{Liske}, {Webb}, {Williger},
  {Fern{\'a}ndez-Soto} \& {Carswell}}{{Liske}
  et~al.}{2000}]{2000MNRAS.311..657L}
{Liske} J.,  {Webb} J.~K.,  {Williger} G.~M.,  {Fern{\'a}ndez-Soto} A.,
  {Carswell} R.~F.,  2000, \mnras, 311, 657

\bibitem[\protect\citeauthoryear{{McDonald} \& {Eisenstein}}{{McDonald} \&
  {Eisenstein}}{2007}]{mcdonald07}
{McDonald} P.,  {Eisenstein} D.~J.,  2007, \prd, 76, 063009

\bibitem[\protect\citeauthoryear{{McDonald} et~al.,}{{McDonald}
  et~al.}{2005}]{mcdonald05b}
{McDonald} P.,  et~al., 2005, \apj, 635, 761

\bibitem[\protect\citeauthoryear{{McDonald}, {Miralda-Escud{\'e}}, {Rauch},
  {Sargent}, {Barlow} \& {Cen}}{{McDonald} et~al.}{2001}]{mcdonald01b}
{McDonald} P.,  {Miralda-Escud{\'e}} J.,  {Rauch} M.,  {Sargent} W.~L.~W.,
  {Barlow} T.~A.,    {Cen} R.,  2001, \apj, 562, 52

\bibitem[\protect\citeauthoryear{{McDonald}, {Miralda-Escud{\'e}}, {Rauch},
  {Sargent}, {Barlow}, {Cen} \& {Ostriker}}{{McDonald}
  et~al.}{2000}]{2000ApJ...543....1M}
{McDonald} P.,  {Miralda-Escud{\'e}} J.,  {Rauch} M.,  {Sargent} W.~L.~W.,
  {Barlow} T.~A.,  {Cen} R.,    {Ostriker} J.~P.,  2000, \apj, 543, 1

\bibitem[\protect\citeauthoryear{{McDonald}, {Seljak}, {Cen}, {Bode} \&
  {Ostriker}}{{McDonald} et~al.}{2005}]{mcdonald05}
{McDonald} P.,  {Seljak} U.,  {Cen} R.,  {Bode} P.,    {Ostriker} J.~P.,  2005,
  \mnras, 360, 1471

\bibitem[\protect\citeauthoryear{{McQuinn}, {Hernquist}, {Lidz} \&
  {Zaldarriaga}}{{McQuinn} et~al.}{2010}]{mcquinn10}
{McQuinn} M.,  {Hernquist} L.,  {Lidz} A.,    {Zaldarriaga} M.,  2010,
  arXiv:1010.5250

\bibitem[\protect\citeauthoryear{{McQuinn}, {Lidz}, {Zaldarriaga}, {Hernquist},
  {Hopkins}, {Dutta} \& {Faucher-Gigu{\`e}re}}{{McQuinn}
  et~al.}{2009}]{mcquinn09}
{McQuinn} M.,  {Lidz} A.,  {Zaldarriaga} M.,  {Hernquist} L.,  {Hopkins} P.~F.,
   {Dutta} S.,    {Faucher-Gigu{\`e}re} C.,  2009, \apj, 694, 842

\bibitem[\protect\citeauthoryear{{McQuinn}, {Oh} \&
  {Faucher-Giguere}}{{McQuinn} et~al.}{2011}]{mcquinn11}
{McQuinn} M.,  {Oh} S.~P.,    {Faucher-Giguere} C.,  2011, arxiv:1101.1964

\bibitem[\protect\citeauthoryear{{Meiksin} \& {White}}{{Meiksin} \&
  {White}}{1999}]{meiksin99}
{Meiksin} A.,  {White} M.,  1999, \mnras, 308, 1179

\bibitem[\protect\citeauthoryear{{Meiksin}}{{Meiksin}}{2009}]{meiksin09}
{Meiksin} A.~A.,  2009, Reviews of Modern Physics, 81, 1405

\bibitem[\protect\citeauthoryear{{O'Meara}, {Prochaska}, {Burles}, {Prochter},
  {Bernstein} \& {Burgess}}{{O'Meara} et~al.}{2007}]{omeara07}
{O'Meara} J.~M.,  {Prochaska} J.~X.,  {Burles} S.,  {Prochter} G.,  {Bernstein}
  R.~A.,    {Burgess} K.~M.,  2007, \apj, 656, 666

\bibitem[\protect\citeauthoryear{{Prochaska}, {O'Meara} \&
  {Worseck}}{{Prochaska} et~al.}{2010}]{prochaska10}
{Prochaska} J.~X.,  {O'Meara} J.~M.,    {Worseck} G.,  2010, \apj, 718, 392

\bibitem[\protect\citeauthoryear{{Reddy} \& {Steidel}}{{Reddy} \&
  {Steidel}}{2009}]{reddy09}
{Reddy} N.~A.,  {Steidel} C.~C.,  2009, \apj, 692, 778

\bibitem[\protect\citeauthoryear{{Rimes} \& {Hamilton}}{{Rimes} \&
  {Hamilton}}{2006}]{RimHam06}
{Rimes} C.~D.,  {Hamilton} A.~J.~S.,  2006, \mnras, 371, 1205

\bibitem[\protect\citeauthoryear{{Schaye}, {Theuns}, {Rauch}, {Efstathiou} \&
  {Sargent}}{{Schaye} et~al.}{2000}]{schaye00}
{Schaye} J.,  {Theuns} T.,  {Rauch} M.,  {Efstathiou} G.,    {Sargent}
  W.~L.~W.,  2000, \mnras, 318, 817

\bibitem[\protect\citeauthoryear{{Seljak} et~al.,}{{Seljak}
  et~al.}{2005}]{2005PhRvD..71j3515S}
{Seljak} U.,  et~al., 2005, \prd, 71, 103515

\bibitem[\protect\citeauthoryear{{Slosar}, {Ho}, {White} \& {Louis}}{{Slosar}
  et~al.}{2009}]{slosar09}
{Slosar} A.,  {Ho} S.,  {White} M.,    {Louis} T.,  2009, Journal of Cosmology
  and Astro-Particle Physics, 10, 19

\bibitem[\protect\citeauthoryear{{Tegmark}, {Taylor} \& {Heavens}}{{Tegmark}
  et~al.}{1997}]{tegmark97}
{Tegmark} M.,  {Taylor} A.~N.,    {Heavens} A.~F.,  1997, \apj, 480, 22

\bibitem[\protect\citeauthoryear{{Viel}, {Lesgourgues}, {Haehnelt}, {Matarrese}
  \& {Riotto}}{{Viel} et~al.}{2005}]{2005PhRvD..71f3534V}
{Viel} M.,  {Lesgourgues} J.,  {Haehnelt} M.~G.,  {Matarrese} S.,    {Riotto}
  A.,  2005, \prd, 71, 063534

\bibitem[\protect\citeauthoryear{{Viel}, {Matarrese}, {Mo}, {Haehnelt} \&
  {Theuns}}{{Viel} et~al.}{2002}]{2002MNRAS.329..848V}
{Viel} M.,  {Matarrese} S.,  {Mo} H.~J.,  {Haehnelt} M.~G.,    {Theuns} T.,
  2002, \mnras, 329, 848

\bibitem[\protect\citeauthoryear{{Weinberg}, {Miralda-Escude}, {Hernquist} \&
  {Katz}}{{Weinberg} et~al.}{1997}]{1997ApJ...490..564W}
{Weinberg} D.~H.,  {Miralda-Escude} J.,  {Hernquist} L.,    {Katz} N.,  1997,
  \apj, 490, 564

\bibitem[\protect\citeauthoryear{{White}}{{White}}{2003}]{white03}
{White} M.,  2003, in The Davis Meeting On Cosmic Inflation {The Ly-a forest}

\bibitem[\protect\citeauthoryear{{White}, {Pope}, {Carlson}, {Heitmann},
  {Habib}, {Fasel}, {Daniel} \& {Lukic}}{{White} et~al.}{2010}]{white10}
{White} M.,  {Pope} A.,  {Carlson} J.,  {Heitmann} K.,  {Habib} S.,  {Fasel}
  P.,  {Daniel} D.,    {Lukic} Z.,  2010, \apj, 713, 383

\bibitem[\protect\citeauthoryear{{Williger}, {Smette}, {Hazard}, {Baldwin} \&
  {McMahon}}{{Williger} et~al.}{2000}]{2000ApJ...532...77W}
{Williger} G.~M.,  {Smette} A.,  {Hazard} C.,  {Baldwin} J.~A.,    {McMahon}
  R.~G.,  2000, \apj, 532, 77

\end{thebibliography}

\appendix

\section{Power-law Luminosity Functions}
\label{ap:powerlaw}

We gain some insight into $\bar{n}_{\rm eff}$
by imagining that the distribution of observed fluxes
for quasars in the target redshift range is a power
law,
\begin{equation}
  N(f)= \frac{N_0}{f_0}
  \left(\frac{f}{f_0}\right)^{-\alpha}
  \qquad (\alpha > 1),
\end{equation}
and that we observe all quasars above a given flux
limit $f_{\rm min}$:
\begin{equation}
  N = N_0 \int_{f_{\rm min}}^\infty \frac{df}{f_0}
      \left(\frac{f}{f_0}\right)^{-\alpha}
    = \frac{N_0}{\alpha-1} 
      \left(\frac{f_{\rm min}}{f_0}\right)^{1-\alpha}.
\end{equation}
If observing conditions are such that the noise
in the forest for a quasar at flux $f_0$ has
$P_{N,n} = \sigma_0^2\,P_{\rm los}$ and scales as
$f^{-2}$ then
\begin{equation}
  \bar{n}_{\rm eff} = n_0
  \int_{f_{\rm min}}^\infty
  \frac{(f/f_0)^{-\alpha}}{1 + \sigma_0^2(f_0/f)^2}
  \ \frac{df}{f_0},
\end{equation}
where $n_0 \equiv  {N_0}/{\mathcal{A}}$.
The behavior of the integral is controlled by
$\Phi\equiv \sigma_0f_0/f_{\rm min}$; for general
$\alpha$ it can be written in terms of a
hypergeometric function, however for integer
$\alpha$ it reduces to elementary functions.
For example, when $\alpha=2$
\begin{equation}
  \bar{n}_{\rm eff} = n_0
  \ \frac{\tan^{-1}\Phi}{\sigma_0},
    \label{eqn:pl2}
\end{equation}
while for $\alpha=3$
\begin{equation}
  \bar{n}_{\rm eff} = n_0
  \ \frac{\ln(1+\Phi^2)}{2\sigma_0^2},
  \label{eqn:pl3}
\end{equation}
and larger $\alpha$ give more negative powers of
$\sigma_0$ and combinations of logs or arctangents.

A natural choice for $f_0$ is to make $\sigma_0=1$,
so that $n_0$ is the number of quasars with
$P_N=P_{\rm los}$ and $\Phi$ measures the minimum
flux in units of the characteristic flux.
It is easy to see that decreasing $f_{\rm min}$,
i.e.~increasing $\Phi$, leads to larger $\bar{n}_{\rm eff}$
but that the gains are small once $\Phi\gg 1$.  For
$f_{\rm min}\sim f_0$ $\bar{n}_{\rm eff}$ is
$n_0$ up to a numerical constant of order
unity, which reinforces the discussion in the text.

\section{Damped Absorbers}
\label{ap:DLAs}

\begin{figure}
\epsfig{file=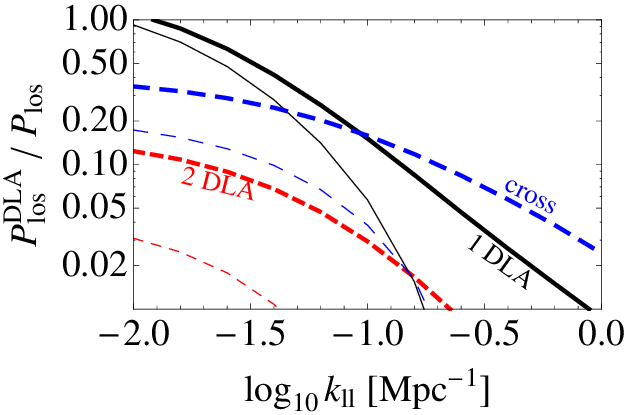, width=8cm}
\epsfig{file=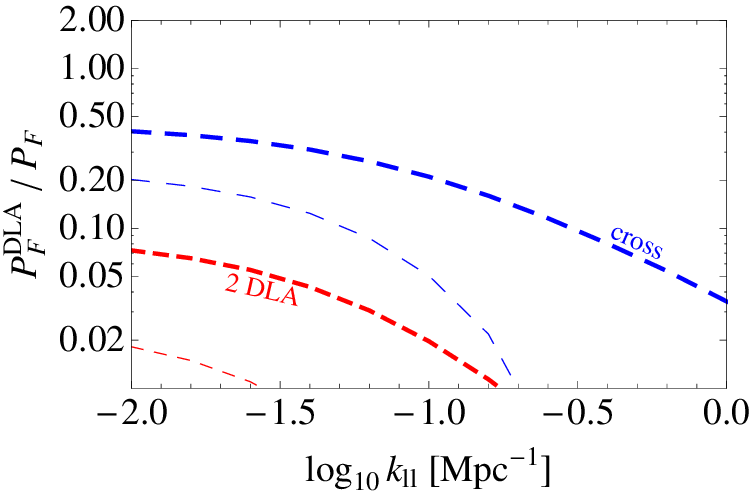, width=8cm}
\caption{Ratio of different components of the power in damping wings to the Ly$\alpha$ forest power without damping wing absorption as a function of $k_{\parallel}$ and at $z=2.5$.  The top panel is for the line-of-sight power spectrum and the bottom is for the $3$D power spectrum.  The thick solid, thick short-dashed, and thick long-dashed curves are respectively the $1$-DLA term, the $2$-DLA term, and the Ly$\alpha$ forest--DLA cross power.  The corresponding thin curves are the same but are calculated only using systems with $N_{\rm HI} > 10^{20}~$cm$^{-2}$.  \label{fig:DLAhalo}}
\end{figure}

As briefly discussed in Section \ref{sec:systematics}, an important systematic for Ly$\alpha$ forest analyses is the damping wing absorption from Lyman-limit and damped Lyman-$\alpha$ systems.  
A simple halo-like model for the contribution to the $3$D power of the absorption from these systems (applicable at scales larger than those affected by thermal broadening) is
\begin{eqnarray}
P_{\rm F}^{\rm DLA} &=&  \widetilde{W}_2(k_\parallel)^2 \, (b_{\rm DLA} + \mu^2)^2 \, P_{\delta}^{\rm lin}(k),\label{eqn:DLAmodel} \\
P_{\rm los}^{\rm DLA} &=&    \widetilde{W}_1(k_\parallel) + \widetilde{W}_2(k_\parallel)^2 \, \int \frac{dk_\perp}{2\pi} k_\perp (b_{\rm DLA} + \mu^2)^2 \, P_{\delta}^{\rm lin}(k),  \nonumber
\end{eqnarray}
where we have neglected the shot noise component in $P_{\rm F}^{\rm DLA}$ because it is small on relevant scales (having amplitude $\sim 1 {\rm ~Mpc}^{-3}$, which is much less than $b_{\rm DLA}^2 \, P_{\delta}^{\rm lin}$) and
\begin{eqnarray}
\widetilde{W}_1(k_\parallel) = \int d N_{\rm HI} \, \frac{\partial^2 {\cal N}}{\partial \chi \partial N_{\rm HI} } \tilde{d}(k_\parallel)^2,\\
\widetilde{W}_2(k_\parallel) = \int d N_{\rm HI} \, \frac{\partial^2 {\cal N}}{\partial \chi \partial N_{\rm HI} } \, \tilde{d}(k_\parallel).
\end{eqnarray}
The function $\tilde{d}$ is the Fourier transform of the damping wing absorption profile, $1 - \exp[-\kappa^2 N_{\rm HI}/(\chi - \chi_0)^2]$, $\chi$ is the conformal distance, and $\chi_0$ is the value of $\chi$ at line-center.  We will refer to the first and second terms in the equation for $P_{\rm los}^{\rm DLA}$ as the $1$-DLA and $2$-DLA terms, respectively.  For details, this is analogous to how the dark matter power spectrum in the halo model is calculated, as reviewed in \citet{cooray02}.  
In addition, $b_{\rm DLA}$ is the bias of such systems (taken to be $2$ here), $\partial^2 {\cal N}/{\partial \chi \partial N_{\rm HI}}$ is the number of systems per $\chi$ per $N_{\rm HI}$, and $\kappa^2 = 5\times 10^{-19}~ {\rm cm}^{-2} {\rm ~Mpc}^{2}$ at $z=3$.  
At small $k_\parallel$, $\widetilde{W}_2$ becomes equal to the fraction of the continuum that is absorbed by damping wings.  Finally, the total effect of damping wings and un-damped Ly$\alpha$ forest absorption on the full $3$D flux power spectrum is
\begin{equation}
P_{\rm F}^{\rm all} = \left[b + b_{\rm DLA} \widetilde{W}_2(k_\parallel) \right]^2 \, P_{\delta}^{\rm lin} + {\rm higher~order}.
\end{equation}


Figure \ref{fig:DLAhalo} plots the fraction of power that originates from damped systems at $z=2.5$.  These calculations assume that $f(N_{\rm HI}, X) \equiv \partial^2 {\cal N}/{\partial z \partial N_{\rm HI}} \sqrt{\Omega_M/(1+z)}$ integrates to $0.1$ between $10^{19}$ and $10^{20.3}~{\rm cm}^{-2}$, is a power-law with index of $-1.2$ at $N_{\rm HI} < 10^{20.3}~{\rm cm}^{-2}$, and with index of $-1.8$ at higher columns, as motivated in \citet{omeara07} and \citet{prochaska10}.  The function $f(N_{\rm HI}, X)$ is uncertain at the factor of $2$-level.  

The top panel in Figure \ref{fig:DLAhalo} shows the ratio of the line-of-sight power in damping wings to $P_{\rm los}$, and the bottom is for the same but for the $3$D power.  The black thick solid, red thick short-dashed, and blue thick long-dashed curves are respectively the $1$-DLA, $2$-DLA terms, and cross-power terms. The $1$-DLA component is only shown in the top panel because it is a subdominant contribution to the $3$D power at the plotted scales owing to the high $3$D number density of these systems.   The $1$-DLA term is a more important contribution than the $2$-DLA to $P_{\rm los}^{\rm all}$, and the cross-power term with the forest, $2 \,b  \,b_{\rm DLA}\, \widetilde{W}_2 P_{\delta}^{\rm lin}$, is important at higher $k_\parallel$.  However, the cross term is the most important contribution to the $3$D forest power (bottom panel).  The thin dashed curves are the same as the thick curves but only include systems with $N_{\rm HI} > 10^{20}~$cm$^{-2}$ (the contribution that is easiest to remove in pre-processing). Systems with $N_{\rm HI} > 10^{20}~$cm$^{-2}$ contribute most of the $1$-DLA power in $P_{\rm los}^{\rm DLA}$ at the smallest $k$, but contribute little of the $1$-DLA power at $k > 0.1~$Mpc$^{-1}$, of the $2$-DLA power in $P_{\rm los}^{\rm all}$ and in $P_{\rm F}^{\rm all}$, or of the cross power.  

The morphology and amplitude of the line-of-sight curves in the top panel of Figure \ref{fig:DLAhalo} are similar to what was found in the numerical calculations of \citet{mcdonald05}, who considered $k_\parallel >0.1~$Mpc$^{-1}$ (see their Fig. 2).  [\citealt{mcdonald05} found a plateau at the highest $k$ shown in Fig. \ref{fig:DLAhalo} that likely owes to the higher order terms, such as $P^{\rm DLA}_{\rm los} \star P_{\rm los}$.]  Our model ignores the coincidence between the normal Ly$\alpha$ absorption and the damping wing absorption:  Damped regions occur where there is already more absorption by un-damped Ly$\alpha$ absorption.  \citet{mcdonald05} found that this correlation suppresses the impact of damping wing absorption by a factor that can be as large as $2$.


Our toy model for the impact of damping wings provides a couple insights.  First, it shows that there is little point of removing the contribution from the highest column systems in $3$D analyses of the forest because they contribute a small component of the total power from damping wings.  Second, because the $1$-DLA term is unimportant in $3$D, this reduces the amplitude of the contribution from damped systems relative to $1$D analyses, especially at the smallest wavevectors.  However, we predict that the contribution from damping wings to $P_{\rm F}$ is still non-negligible.  Lastly, in a $3$D survey one has the ability to fit for the the damping wing contribution:  It should scale at relevant scales as $\mathcal{C}(k_\parallel) \, P_{\delta}^{\rm lin}(\bfk)$, where $\mathcal{C}$ is some function that only depends on $k_\parallel$.

\end{document}